\documentclass[12pt]{iopart}

\usepackage{graphicx}
\begin{document}

\title[Vacuum properties]{Magnetic and electric properties of quantum vacuum}

\author{R. Battesti, C. Rizzo}

\address{Laboratoire National des Champs Magn\'etiques Intenses, 143 avenue de Rangueil, 31400 Toulouse, France.}
\ead{remy.battesti@lncmi.cnrs.fr}
\begin{abstract}
In this report we show that vacuum is a nonlinear optical medium and we discuss what are the optical phenomena that should exist in the framework of the standard model of particle physics. We pay special attention to the low energy limit. The predicted effects for photons of energy smaller than the electron rest mass are of such a level that none has been observed experimentally yet. Progresses in field sources and related techniques seem to indicate that in few years vacuum nonlinear optics will be accessible to human investigation.
\end{abstract}

\maketitle

\tableofcontents

\section{Introduction}

Since very ancient times the existence of the vacuum is one of the most fundamental problems in science.
In the Aristotle's Physics one reads that "the investigation of questions about the vacuum must be held to belong to the
physicist - namely whatever it exists or not, and how it exists or what it is"\cite{Aristotle}. In his work, Aristotle defines what is place, time, and vacuum which "must, if it exists, be place deprived of body"\cite{Aristotle}.

Following the cyclic evolutions of the concept of vacuum in the history of physics is out of the scope of this report, but we have at least to define what we intend for vacuum in the following. We are interested in the electromagnetic properties of vacuum, so our definition is deeply related to electromagnetism : a vacuum is a region of space in which a monochromatic electromagnetic plane wave propagates at a velocity that is equal to $c$. Following special relativity, this velocity is independent both on the source motion and on the observer inertial frame of reference. Such a vacuum has not to be empty. For example in XIX century classical electrodynamics vacuum was supposed to be filled by the ether. Our definition is essentially a phenomenological one. In principle it provides a way to test if a region is a vacuum or not.

In classical electrodynamics, vacuum electromagnetic properties are simply represented by two fundamental constants :  the vacuum permittivity $\epsilon_0$ and the vacuum permeability $\mu_{0}$. These constants are linked to the velocity of light in vacuum $c$ thanks to the fundamental relation $c={{1}\over{\sqrt{\epsilon_0 \mu_0}}}$. They describe respectively the proportionality factor between $D$ and $E$ and between $B$ and $H$ in vacuum: \textbf D=$\epsilon_0$\textbf E and \textbf B=$\mu_0$\textbf  H, where \textbf D is the electric displacement vector, \textbf E is the electric field vector, \textbf B is the magnetic induction vector and \textbf  H is the magnetic field vector. \textbf B is also called magnetic field vector as well as \textbf  H when there is no risk of confusion \cite{Jackson}. We conform to this common terminology in the following.

Any variation of the velocity of light with respect to $c$ is ascribed to the fact that light is propagating in a medium, i.e. not in vacuum. To describe such a phenomenon one introduces the constants $\epsilon$ and $\mu$ which characterize the medium itself: \textbf D=$[\epsilon]$ \textbf E and \textbf B=$[\mu]$ \textbf H. The velocity of light in a medium is smaller than the velocity of light in vacuum by a factor $n$, the index of refraction, equal to $n={{\sqrt{\epsilon \mu}}\over{\sqrt{\epsilon_0 \mu_0}}}$. Vacuum is therefore the medium to which in classical electrodynamics one associates an index of refraction $n$ exactly equal to 1 and therefore in vacuum ${\epsilon \over \epsilon_0}=1$ and ${\mu \over \mu_0}=1$.

Since the half of XIX century, when Faraday discovered that an external magnetic field could change the polarization of light propagating in matter because of a magnetically induced circular birefringence \cite{Faraday1845}, it is known that the presence of electrostatic fields induces a response from the medium which depends on the field strength. This means that $\epsilon$ and $\mu$ are not constants but functions of the external fields. In 1961 the second harmonic generation experiment of Franken et al. \cite{Franken1961} opened up the field of nonlinear optics. In general, it deals with nonlinear interactions of light with matter including light and/or external electromagnetic field induced changes of the optical properties of a medium \cite{Shen}. All media are basically nonlinear and thus $\epsilon$ and $\mu$ can be written as $\epsilon$(\textbf{E,B}) and $\mu$(\textbf{E,B}) where \textbf{E} and \textbf{B} are the total electromagnetic fields both external and associated to the electromagnetic waves propagating in the medium.  Therefore in general $n$ depends also on \textbf{E} and \textbf{B}, $n$(\textbf{E,B}).

In this report we show that vacuum can be considered as a nonlinear optical medium and we discuss what are the optical phenomena that should exist in the framework of the standard model of particle physics. Our work is mainly devoted to the phenomenological manifestations of these quantum vacuum non linearities. We classify all the expected phenomena following nonlinear optics textbooks. We pay special attention to the low energy limit since this limit is sufficient to understand most of the attempted and proposed experiments. While the corresponding theory is almost a century old, the predicted effects for photons of energy smaller than the electron rest mass are of such a level that none has been observed experimentally yet. Progresses in field sources and related techniques seem to indicate that in few years vacuum nonlinear optics will be accessible to human investigation.

\section{Theory}

\subsection{General formalism}
In a medium the excitation due to light and external fields produces a polarization \textbf P and a magnetization  \textbf M. Both of them depend on the electromagnetic fields and hence the response of the medium to the excitation is nonlinear.
For describing this nonlinear interaction, one uses the constitutive equations of the medium giving the relationship between \textbf P and (\textbf E,\textbf B)
and between \textbf M and (\textbf E,\textbf B), and
Maxwell's equations \cite{Shen}.
When no charge density or current density are present, Maxwell's equations can be written in SI units \cite{Jackson} as

\begin{eqnarray}
 \nonumber \nabla \times \textbf{E} &=& -\frac{\partial \textbf{B}}{\partial t}, \\
  \nabla \times \textbf{H} &=& \frac{\partial \textbf{D}}{\partial t},\\
 \nonumber \nabla\cdot\textbf{D}    &=& 0, \\
 \nonumber \nabla\cdot \textbf{B}   &=& 0,
\end{eqnarray}

with

\begin{eqnarray}\label{Con}
\nonumber \textbf{H} &=& \frac{1 }{\mu_0}\textbf{B}-\textbf{M},\\
\textbf{D} &=& {\epsilon_0}\textbf{E}+\textbf{P}.
\end{eqnarray}

Thanks to Maxwell's equations one can fully determine wave propagation.

The constitutive equations can be obtained by the following relations \cite{Landau}:

\begin{equation}
\label{Di}
\textbf{D} = {\partial L \over \partial \textbf{E}},
    \end{equation}

\begin{equation}
\label{acca}
\textbf{H} = -{\partial L \over \partial \textbf{B}},
    \end{equation}

where $L$ is the effective lagrangian representing the interaction of electromagnetic fields in vacuum.

The mathematical expression of the effective lagrangian $L$ is essentially determined by the fact that it has to be relativistic invariant and therefore it can only
be a function of the Lorentz invariants  F and G \cite{Landau} :
\begin{equation}
F =  \left(\epsilon_{0}E^2 - {B^2 \over \mu_{0}}\right),
    \label{2}
    \end{equation}
\begin{equation}
G =\sqrt{\epsilon_{0} \over \mu_{0}} (\textbf{E} \cdot \textbf{B}).
    \label{3}
    \end{equation}

The general expression can be therefore written as

\begin{equation}\label{genform}
    L= \sum_{i=0}^\infty \sum_{j=0}^\infty c_{i,j} F^i
    G^{j}.
\end{equation}

The lowest order terms must give the classical Maxwell lagrangian $L_0=\frac{1}{2}F$, thus $c_{0,0}=0$, $c_{1,0}=\frac{1}{2}$ and  $c_{0,1}=0$.


Using the relations (\ref{Di}) and (\ref{acca}) one obtains

\begin{eqnarray}\label{Digenform}
    \nonumber \textbf{D} &=& \sum_{i=0}^\infty \sum_{j=0}^\infty c_{i,j} \left(iF^{(i-1)}G^{j}{\partial F \over \partial \textbf{E}} + j
    G^{(j-1)}F^i{\partial G \over \partial \textbf{E}}\right)\\
    &=&\sum_{i=0}^\infty \sum_{j=0}^\infty c_{i,j} \left(2\epsilon_0iF^{(i-1)}G^{j}\textbf{E} + j\sqrt{{\epsilon_{0}
\over \mu_{0}} }
    G^{(j-1)}F^i\textbf{B}\right),
\end{eqnarray}

\begin{eqnarray}\label{accagenform}
    \nonumber \textbf{H} &=& \sum_{i=0}^\infty \sum_{j=0}^\infty -c_{i,j} \left(iF^{(i-1)}G^{j}{\partial F \over \partial \textbf{B}} + j
    G^{(j-1)}F^i{\partial G \over \partial \textbf{B}}\right)\\
    &=&\sum_{i=0}^\infty \sum_{j=0}^\infty c_{i,j} \left(2iF^{(i-1)}G^{j}{\textbf{B} \over \mu_0} - j\sqrt{{\epsilon_{0}
\over \mu_{0}} }
    G^{(j-1)}F^i\textbf{E}\right),
\end{eqnarray}

which at the lowest orders gives

\begin{eqnarray}\label{Digen}
\nonumber
\textbf{D} &=
& 2\epsilon_{0}{c_{1,0}}\textbf{E} + \sqrt{{\epsilon_{0}
\over \mu_{0}} }{c_{0,1}}\textbf{B} +2\epsilon_{0}{c_{1,1}}G\textbf{E} +\sqrt{{\epsilon_{0}
\over \mu_{0}} }{c_{1,1}}F\textbf{B}
+4\epsilon_{0}{c_{2,0}}F
\textbf{E}\\
& &+2 \sqrt{{\epsilon_{0}
\over \mu_{0}} }{c_{0,2}}G\textbf{B},
\end{eqnarray}

\begin{eqnarray}\label{accagen}
\nonumber
\textbf{H} &=
& 2{c_{1,0}}{\textbf{B} \over \mu_{0}} - \sqrt{{\epsilon_{0}
\over \mu_{0}} }{c_{0,1}}\textbf{E} +2{c_{1,1}}G{\textbf{B} \over \mu_{0}} -\sqrt{{\epsilon_{0}
\over \mu_{0}} }{c_{1,1}}F\textbf{E}
+4{c_{2,0}}F
{\textbf{B} \over \mu_{0}}\\
& &-2 \sqrt{{\epsilon_{0}
\over \mu_{0}} }{c_{0,2}}G\textbf{E}.
\end{eqnarray}

The classical equations ${\bf D}=\epsilon_0 {\bf E}$ and ${\bf H} = \frac{\bf B}{\mu_0}$ are recovered at the lowest order in the fields by imposing $c_{1,0}=\frac{1}{2}$ and  $c_{0,1}=0$.

In the case of a plane wave propagating in a vacuum, both $F$ and $G$ are equal
to zero and therefore $L = 0$ as well. This means that, because of Lorentz invariance, the propagation of a plane wave in vacuum cannot be affected by any nonlinear interactions. It can be shown that $L$ is also equal to 0 in the case of two copropagating (${\bf k_1}={\bf k_2}$) plane waves of different polarization (${\bf E_1}\neq{\bf E_2}$). The simplest cases in which $L \neq 0$, giving rise to nonlinear effects in vacuum, are the one of a plane wave propagating in the presence of external static electric or magnetic fields (${\bf E_0}$, ${\bf B_0}$), and the one of two plane waves of the same polarization (${\bf E_1}={\bf E_2}$) of different wavevectors (${\bf k_1}\neq{\bf k_2}$). This is true in particular for counterpropagating plane waves (${\bf k_1}={-\bf k_2}$).

It is worth stressing that, as far as Lorentz invariance holds, our phenomenological definition of a vacuum also holds even in the presence of nonlinear interactions. A plane wave velocity different from $c$ can be ascribed to the presence of matter, as in classical electrodynamics, and/or to the presence of electromagnetic fields.

Since $L$ can only be a sum of terms containing powers of $F$ and $G$, lagrangian terms containing a product of an odd number of electromagnetic fields are not allowed in vacuum (see eq. \ref{genform}). This means that not all the nonlinear effects existing in a standard medium exist in vacuum. The form of lagrangian $L$ also indicates that the $B$ field and the $E$ field of waves play an equivalent role as far as nonlinear effects in vacuum are concerned. In standard media $B$ field is usually neglected and all the effects are ascribed to an $\epsilon$ function only of $E$ while $\mu$ is assumed equal to $\mu_0$ \cite{Shen}. In this sense a vacuum can be considered as a magnetic medium.

Finally, the energy density $U$ can be written as \cite{Landau}

\begin{eqnarray}\label{Ener}
\nonumber U &=& \textbf{E} {\partial L \over \partial \textbf{E}} - L\\
&=& \sum_{i=0}^\infty \sum_{j=0}^\infty c_{i,j} \left(2\epsilon_0iF^{(i-1)}G^{j}E^2 + (j-1)
    F^iG^{j}\right).
\end{eqnarray}

Taking into account that $c_{0,0}=0$, $c_{1,0}=\frac{1}{2}$ and  $c_{0,1}=0$, the term $(i=1,j=0)$ gives the classical energy density $U_0$.

\begin{equation}\label{Umax}
    U_0=\frac{1}{2}\left(\epsilon_0 E^2 + {B_0^2 \over \mu_0}\right).
\end{equation}

This expansion of the electromagnetic energy density in vacuum can be compared with the one given by Buckingham \cite{Buckingham} in the case of a molecule in the presence of external electromagnetic fields. One can find a direct correspondence between vacuum terms and the molecular ones, showing once more that vacuum behaves as a standard medium. In the case of molecular energy density each term represents a specific microscopical property of a single molecule, like polarizability, while for vacuum we are dealing with macroscopical properties. To pass from microscopical to macroscopical properties in the case of molecules one has to take into account the molecular density, a concept that has no equivalence for vacuum.

In the different theoretical frameworks one can find predictions for the $c_{i,j}$ coefficients introduced in Eq. (\ref{genform}). Quantum electrodynamics (QED) provides the most complete theoretical treatment.

\subsection{QED effective lagrangian}

In 1933 the calculations of gamma rays absorption due to the formation of electron-positron pairs by Oppenheimer and Plesset \cite{Oppenheimer1933} gave a striking confirmation of Dirac's theory of the positron as holes in a sea of negative energy states setting up a new picture of the vacuum in which all the negative energy states are occupied and all the positive energy states are unoccupied. Dirac gives a clear overview of his model in his 1934 contribution to the Solvay workshop \cite{Dirac1934}. It was immediately clear that an important prediction of Dirac's theory, which could in principle be experimentally tested, was the existence of photon-photon scattering \cite{Halpern1933}, \cite{Breit1934}.

A first theoretical formulation of optical non linearities in vacuum at the lowest orders in the electromagnetic fields has been published in 1935 by Euler and Kochel \cite{Euler1935}.  The details about their calculation can be found in ref. \cite{Euler1936}. In the 1936 paper by Heisenberg and Euler \cite{Heisenberg1936} a complete theoretical study of the phenomena related to the fact that electromagnetic radiation can be transformed into matter and vice versa can be found. The authors starting point was that it was no more possible to separate processes in the vacuum from those involving matter since electromagnetic fields can create matter if they are strong enough. Moreover, even if they are not strong enough to create matter, they polarize the vacuum because of the virtual possibility of creating matter, essentially with electron-positron pairs, and therefore they change the constitutive equations \cite{Heisenberg1936}.

The resulting effective lagrangian of the field reads \cite{Heisenberg1936}:

\begin{eqnarray}\label{LHE}
L_{HE} = {1 \over 2}\left(\epsilon_{0}E^2 - {B^2 \over \mu_{0}}\right)+\alpha\int_0^\infty e^{-\eta}{d\eta \over \eta^3}& & \\  \nonumber \times \left\lbrace  i\eta^2\sqrt{\epsilon_{0} \over \mu_{0}} (\textbf{E} \cdot \textbf{B})  \frac{cos\left( \frac{\eta}{\sqrt{\epsilon_{0}}E_{cr}}\sqrt{C}\right)+ conj.}{cos\left( \frac{\eta}{\sqrt{\epsilon_{0}}E_{cr}}\sqrt{C}\right)- conj.} + \epsilon_{0}E_{cr}^2+\frac{\eta^2}{3}(\epsilon_{0}E^2 - {B^2 \over \mu_{0}})\right\rbrace, & &\\
\nonumber \textrm{with } C =\left(\epsilon_{0}E^2 - {B^2 \over \mu_{0}}\right)+2i{\epsilon_{0} \over \mu_{0}} (\textbf{E} \cdot \textbf{B}),& &
\end{eqnarray}

where $\alpha = {e^2 \over 4\pi\epsilon_0\hbar c}$ is the fine structure constant, $e$ the elementary charge, $\hbar$ the Planck constant $h$ divided by $2\pi$, and $\eta$ is the integration variable.
$E_{cr} = {m_e^2c^3 \over e\hbar}$ is a quantity obtained by combining the fundamental constant $m_e$, the electron mass, with $c$, $e$ and $\hbar$. $E_{cr}$ has the dimensions of an electric field, and it is called the critical electric field. Its value is $E_{cr} = 1.3 \times 10^{18}$ V/m. It corresponds to the field one needs to get an energy $eE_{cr}L$ equal to an electron rest mass $m_ec^2$ over a length $L$ equal to the reduced electron Compton wavelength $\mathchar'26\mkern-10mu\lambda = {\lambda_e \over 2\pi} = {\hbar \over m_ec}$.

A critical magnetic field can also be defined in the same manner, $B_{cr} = {E_{cr} \over c} = {m_e^2c^2 \over e\hbar}$. $B_{cr} = 4.4 \times 10^9$ T. The cyclotron pulsation for an electron in a $B_{cr}$ field, $\omega_c = {eB_{cr} \over m}$, is such that the associated energy $\hbar \omega_c$ is equal to its rest mass $m_ec^2$.

$L_{HE}$ is valid in the approximation that the
fields vary very slowly over a length equal to the reduced
electron Compton wavelength during a time
${{t_{e}}={\mathchar'26\mkern-10mu\lambda \over c}}$ which corresponds to:

\begin{equation}
{\hbar \over m_{e}c} {|\nabla E (B)|} \ll E (B),
    \end{equation}
\begin{equation}
{\hbar \over m_{e}c^2} |{\partial E (B)\over \partial t} | \ll E
(B).
    \end{equation}

A study of the vacuum electrodynamics based on the quantum theory of the electron
can also be found in the 1936 paper by Weisskopf \cite{Weisskopf1936} in which a simplified method to obtain $L_{HE}$ is shown.

In general QED lagrangian $L_{HE}$ can be expanded as indicated in Eq. (\ref{genform}).
Thanks to the symmetry properties of the ${\bf E}$ and ${\bf B}$ fields (see table 1), $F$ and $G$ are also CPT invariant, but while $F$ is C, P, and T invariant, $G$ violates P and T.

\begin{table}[htdp]
\begin{center}\begin{tabular}{|c|c|c|c|c|}\hline  & C & P & T & CPT \\\hline \bf E & - & - & + & + \\\hline \bf B & - & + & - & + \\\hline F & + & + & +  & + \\\hline G & + & - & - & + \\\hline \end{tabular}
\caption{Symetry properties on the electromagnetic fields}
\end{center}
\label{symCha}
\end{table}

As the QED quantum vacuum is assumed to be C, P, T invariant, one has to take into account that all the  coefficients $c_{i,j}$ with an index $j$ corresponding to an odd number are null, and in particular $c_{0,1}=0$ and $c_{1,1}=0$.

At the lowest orders in the fields, i.e. $E \ll E_{cr}$ and $B \ll B_{cr}$, $L_{HE}$ can be written as
$L_{HE} = L_{0} + L_{EK}$ where $L_{EK}$ has been first
calculated by Euler and Kockel in 1935 \cite{Euler1935}. $L_{EK}$ can be written as $L_{EK} = c_{2,0}F^2 + c_{0,2}G^2$

The value of $c_{2,0}$ and $c_{0,2}$ can be written following the
Euler-Kochel result \cite{Euler1935} as :

\begin{equation}
c_{2,0} = {2\alpha^2 \hbar^3 \over 45 m_{e}^4 c^5} = {\alpha  \over 90 \pi} {1 \over \epsilon_0 E_{cr}^2} = {\alpha  \over 90 \pi} {\mu_0 \over B_{cr}^2} \simeq 1.67 \times 10^{-30}~\left[{m^3 \over J}\right],
\end{equation}

\begin{equation}
c_{0,2} = 7 c_{2,0},
\end{equation}

and therefore

\begin{equation}
L_{EK} = {2\alpha^2 \hbar^3 \over 45 m_{e}^4 c^5} \epsilon_0^2[ (E^2-c^2B^2)^2+7c^2(\textbf E\cdot \textbf B)^2].
\label{LEK}
\end{equation}

Next term coefficients can be found in  \cite{Heisenberg1936}, \cite{Weisskopf1936} and \cite{Bialynicka-Birula1970}.

\begin{equation}
c_{3,0} = {32\pi \alpha^3 \hbar^6 \over 315 m_{e}^8 c^{10}} = {2\alpha  \over 315 \pi} {1 \over \epsilon_0^2 E_{cr}^4} = {2\alpha  \over 315 \pi} {\mu_0^2 \over B_{cr}^4} \simeq 6.2 \times 10^{-56}~\left[{m^6 \over J^2}\right],
\end{equation}

\begin{equation}
c_{1,2} = {13 \over 2} c_{3,0},
\end{equation}

\begin{equation}
c_{4,0} = {3568\pi^2 \alpha^4 \hbar^9 \over 945 m_{e}^{12} c^{15}} = {223\alpha  \over 3780 \pi} {1 \over \epsilon_0^3 E_{cr}^6} = {223\alpha  \over 3780 \pi} {\mu_0^3 \over B_{cr}^6} \simeq 4.4 \times 10^{-80}~\left[{m^9 \over J^3}\right],
\end{equation}

\begin{equation}
c_{2,2} = {402 \over 223}c_{4,0},
\end{equation}

\begin{equation}
c_{0,4} = {304 \over 223}c_{4,0}.
\end{equation}

Following the developments in QED of the forties by Tomonaga, Schwinger, Feynman and Dyson \cite{Libro}, a quantum field theory has been fully established and the Dirac's model of vacuum has become somewhat an obsolete and unnecessary concept. In the field theory perspective, vacuum becomes the ground state of the quantum field. This theoretical definition of a vacuum is in agreement with ours which refers to macroscopical phenomenological electromagnetic properties like $\epsilon$ and $\mu$ \cite{Jauch1948}.

\begin{figure}[ht]
\begin{center}
  \includegraphics[width=4cm]{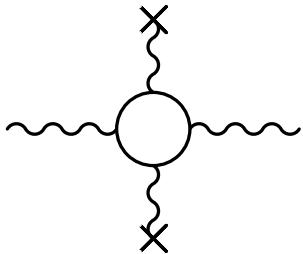}
  \end{center}
  \caption{Feynman diagram for the lowest order in the nonlinear effect in vacuum. Solid circular lines represent the electron-positron loops, wavy lines the photons and wavy lines with cross ending the external fields.}\label{FD}
\end{figure}

A modern confirmation of Euler and Kochel result has been given by Karplus and Neuman in 1950 \cite{Karplus1950}, and the Heisenberg-Euler lagrangian has been validate by Schwinger in 1951 \cite{Schwinger1951}. The lowest order in the nonlinear effect given by the Euler-Kochel lagrangian can be represented by the Feynman diagram of Fig. \ref{FD} \cite{Karplus1950}.

An analytic form for the Heisenberg-Euler lagrangian can be found in ref. \cite{Heyl1997}, but for most of the practical cases, effects can be calculated using first terms given by $L_{EK}$. In fact eq. (\ref{LHE}) can be expanded in terms of reduced Lorentz invariants $F'$ and $G'$

\begin{equation}
F' =  \left(\frac{E^2}{E_{cr}^2} - {B^2 \over B_{cr}^2}\right),
    \label{2red}
    \end{equation}
\begin{equation}
G' =\left(\frac{\textbf{E}}{E_{cr}} \cdot \frac{\textbf{B}}{B_{cr}}\right),
    \label{3red}
    \end{equation}

which means that higher and higher order terms corresponding to higher and higher powers of $F'$ and $G'$ give smaller and smaller contributions when $\frac{E}{E_{cr}}$ and $\frac{B}{B_{cr}}$ are much smaller than 1.

After 75 years the impact of Heisenberg-Euler lagrangian in fundamental physics is still very important as discussed in ref. \cite{Dunne2012}. Our review is essentially based on its predictions.

\begin{figure}[h]
\begin{center}
  \includegraphics[width=4cm]{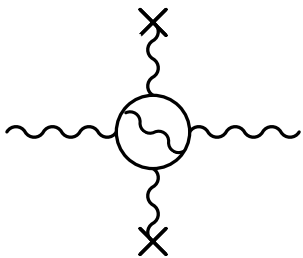}
  \end{center}
  \caption{The Feynman diagram corresponding to the lowest order radiative corrections.}\label{FD2}

\end{figure}

Heisenberg-Euler lagrangian does not take into account all the microscopic phenomena related to the photon-photon interaction in vacuum. Corrections to the value of the coefficients $c_{i,j}$ obtained using $L_{HE}$ can be calculated taking into account the change induced by the external fields in the radiative interactions of the vacuum electrons. In particular Ritus \cite{Ritus1975} has published in 1975 the corrections to $c_{2,0}$ and $c_{0,2}$.
The lowest order radiative corrections can be represented by the Feynman diagram of Fig. \ref{FD2},

and the corresponding effective lagrangian can be written following \cite{Ritus1975} as

\begin{equation}
L_R =  \frac{\alpha^3 \hbar^3}{81\pi m_{e}^4 c^5}\left(16F^2+\frac{263}{2}G^2\right),
    \label{Lritus}
    \end{equation}

which gives a 1.0\% correction to $c_{2,0}$:

\begin{equation}
c_{2,0}^R = {2\alpha^2 \hbar^3 \over 45 m_{e}^4 c^5} \left(1 + {40 \alpha \over 9\pi} \right)
\end{equation}

and a 1.2\% correction to $c_{0,2}$:

\begin{equation}
c_{0,2}^R = {14\alpha^2 \hbar^3 \over 45 m_{e}^4 c^5} \left(1 + {1315 \alpha \over 252\pi} \right).
\end{equation}

The Ritus corrections to $c_{2,0}$ and $c_{0,2}$ are about $\alpha$ times smaller than the Euler-Kochel values for these two coefficients. Nevertheless, these corrections are more important than the next terms in the expansion of $L_{HE}$ corresponding to the coefficients $c_{1,2}$, $c_{3,0}$ when $E \ll E_{cr}$ and $B \ll B_{cr}$.

Following Eq. (\ref{Ener}), the energy density $U$ when $E \ll E_{cr}$ and $B \ll B_{cr}$ can be written as \cite{Weisskopf1936}:

\begin{eqnarray}\label{Ude}
\nonumber  U &=&{1 \over 2}\left(\epsilon_0 E^2 + {B^2 \over \mu_{0}}\right) + c_{2,0}\left(\epsilon_0 E^2 - {B^2 \over \mu_{0}}\right)
     \left(3\epsilon_0 E^2 + {B^2 \over \mu_{0}}\right) \\
   &+&c_{0,2}{\epsilon_{0} \over \mu_{0}} (\textbf{E} \cdot \textbf{B})^2 + c_{3,0}\left(\epsilon_0 E^2 - {B^2 \over \mu_{0}}\right)^2\left(5\epsilon_0 E^2 + {B^2 \over \mu_{0}}\right)\\
\nonumber   &+& c_{1,2}{\epsilon_{0} \over \mu_{0}} (\textbf{E} \cdot \textbf{B})^2\left(3\epsilon_0 E^2 - {B^2 \over \mu_{0}}\right).
\end{eqnarray}

Using the relations (\ref{Digen}), (\ref{accagen}) and (\ref{Con}) one obtains the polarization and the magnetization of the vacuum at the lowest orders in the fields :

\begin{equation}\label{Pol}
\textbf{P} = 4c_{2,0}
\epsilon_{0}\textbf{E}F + 2c_{0,2} \sqrt{{\epsilon_{0}
\over \mu_{0}} }\textbf{B}G,
\end{equation}

\begin{equation}\label{Magn}
\textbf{M} = -
4c_{2,0} {\textbf{B} \over \mu_{0}}F + 2c_{0,2} \sqrt{{\epsilon_{0} \over \mu_{0}} }\textbf{E}G.
\end{equation}

In principle, since electromagnetic fields have self-interactions one should calculate the corrections to the Maxwell classical solutions for any distribution of charges and currents \cite{Serber1935}.

The corrections to the Coulomb potential of a charge induced by vacuum polarization have been calculated in 1935 by Uehling \cite{Uehling1935} together with the corresponding displacement of atomic energy levels. This is a fundamental result for the QED of bound states.

The case of a magnetic dipole has been treated in ref. \cite{Heyl1997_2} where the field equations of a static magnetic field have been considered and the field of the dipole has been calculated taking into account one-loop QED corrections.

The value of the lowest order coefficients $c_{2,0}$ and $c_{0,2}$ depends on the fourth power of the inverse of the electron mass $m_e$. Following ref. \cite{Schwinger1951} one can generalize Heisenberg-Euler result to any spin ${1 \over 2}$ charged field like the one corresponding to negative and positive muons. For the sake of comparison, one can consider that the value of $B_{cr}$ sets the relative scale of different contributions coming from different fermions. The critical magnetic field for muon leptons is about 1.9 $\times$ 10$^{14}$ T and for tau leptons 5.3 $\times$ 10$^{16}$ T. The contributions coming from muon or tau fields are usually neglected.

\subsection{Other contributions to the effective lagrangian $L$}

The effective lagrangian $L$ is valid in the approximation of constant or slowly varying electromagnetic fields. As the available laser pulse intensity increases and pulse time width decreases, one needs in principle to take into account the dispersive corrections of $L$. This means that one has to take into consideration a correction to $L$ depending on the derivatives of electromagnetic fields. How to treat dispersion and absorption in the lagrangian formalism is a debated question and it has generated a large literature to correctly describe quantum electromagnetism in dielectric media. From this literature it is worth mentioning the two recent works of Huttner and Barnett \cite{Huttner1992}, and Philbin \cite{Philbin2010}.

As far as quantum vacuum is concerned, the authors of ref. \cite{Shukla2004} give the effective lagrangian corresponding to dispersion corrections and they derive in the low energy limit ($\hbar \omega \ll m_e c^2$) the vacuum dispersion relation:

\begin{equation}\label{DisRel}
\omega \simeq c k \left(1 - \frac{1}{2}\zeta Q^2 - \sigma \zeta^2 Q^4 k^2 \right),
\end{equation}

where $\zeta$ depends on light polarization and it is equal to $4c_{2,0}$ or $2c_{0,2}$, $\sigma = \frac{2\alpha\hbar^2}{15m_e^2 c^2}$ is a parameter corresponding to the dispersive properties of the polarized vacuum. Its numerical value is about $1.4$ $\times$ 10$^{-28}$ m$^2$ and $Q$ is a parameter depending on the electromagnetic fields. For electromagnetic fields perpendicular to the light wavevector $k$, $Q$ can be written as

\begin{equation}\label{Qu}
Q^2 = \epsilon_0 E^2 + \frac{B^2}{\mu_0}-2\sqrt{\frac{\epsilon_0}{\mu_0}}(\textbf{E}\times\textbf{B}).
\end{equation}

The first two terms of eq. (\ref{DisRel}) correspond to a linear dispersion relation representing a vacuum velocity of light that depends slightly on the light polarization. This is an important point that will be treated in details in the following paragraphs.  The third term is the nonlinear term coming from the dispersive correction to $L$. Its order of magnitude can be estimated as $8 \times 10^{-86} Q^4 k^2$, and it seems very challenging to detect in a laboratory \cite{Shukla2004}.

In the framework of the standard model, contributions others than QED ones appears essentially at the QCD scale and at the electroweak scale. QCD scale can be associated to a mass $\Lambda_{QCD} \simeq 200$ MeV  close to the mass of the pion $\pi^\pm$ meson. The corresponding critical magnetic field $B^{QCD}_{cr}$ is of the order of 10$^{15}$ T \cite{Kabat2002}. Electroweak scale can be associated to the mass of the $W^\pm$ boson which is about 80 GeV . The corresponding critical magnetic field $B^{EW}_{cr}$ is of the order of 10$^{20}$ T\cite{Kabat2002}.
It is obvious that these contributions can be neglected most of the time and there is not much literature about them.

\section{Instruments}

The experimental tests of the Heisenberg-Euler lagrangian need high magnetic or electric fields. In this section, we give a short overview of the existing solutions for producing such intense fields.

\subsection{Light sources}

Since 1960 and the invention of the laser by Maiman \cite{Maiman}, the available intensity has jumped to highest and highest levels. Today, a number of exawatt ($10^{18}$ W) class facilities are already in the planning stage (ELI program in Europe \cite{ELI}, Exawatt Laser in Japan \cite{ExaJapan} for example).
High power lasers are expected to approach in the future the Schwinger limit corresponding to an intensity $I_s$ $\approx$ 10$^{33}$ W.m$^{-2}$. At such a level the electric field $E_\omega$ associated to the wave is of the order of the critical one, $E_\omega = \sqrt{\frac{I_s}{2\epsilon_0 c}} \approx E_{cr}$, which allows the creation of real $e^-e^+$ pairs from the vacuum \cite{Schwinger1951}.

All facilities around the world are nowadays based on the Chirped Pulse Amplification (CPA) proposed by G. Mourou in 1985 \cite{Mourou}. Since then the available laser power has increased rapidly. Basically, high powers are limited by the damage threshold of materials which typically lies around 10$^4$ GW.m$^{-2}$. The problem thus is to amplify a short pulse without reaching this damage intensity. The principle of the CPA is the following (see fig. (3)) : a short laser impulsion  is temporally stretched before being amplify up to 10$^4$ GW.m$^{-2}$. Then the pulse is temporally recompressed  before being focused on a target. In this way, one can amplify the intensity (expressed in J.s$^{-1}$.m$^{-2}$) of the pulse without decreasing its fluency (expressed in J.m$^{-2}$).

The Extreme Light Infrastructure (ELI) project represents one of the largest laser project in the world. Its goal is to produce a few kJ of energy in 10 fs, which means more than 10$^{14}$ W of power with a target intensity of 10$^{30}$ W.m$^{-2}$ \cite{ELI}. As far as we know, the highest intensity ever reached nowadays is 2 $\times$ 10$^{26}$ W.m$^{-2}$ \cite{Yanovsky2008} with the HERCULES petawatt facility \cite{HERCULES} in the USA.

Two ambitious projects are also in progress, the National Ignition Facility, USA \cite{NIF} and the laser Megajoule, France \cite{MegaJ}, in order to create fusion ignition in laboratory. Both of them will be able to fire an energy of about 1,8 MJ thanks to the amplification of more than 200 laser beams.

\begin{figure}[h]
\begin{center}
  \includegraphics[width=10cm]{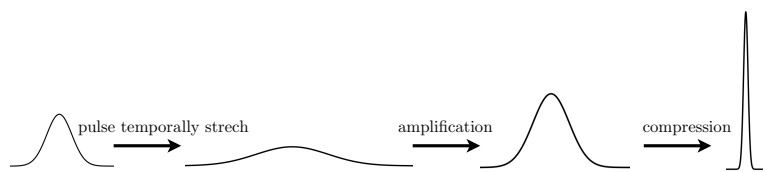}
\label{CPA}
    \end{center}
      \caption{Chirped Pulse Amplification technique (CPA). A short laser pulse is temporally streched. After this operation the fluency (J/m$^2$) is the same as before but the intensity is reduced. Then the pulse can be amplified without risk for the optical materials.  When the desired energy is achieved, the pulse is recompressed to reach high intensities. }
\end{figure}

For smaller intensities, commercial sources are also available. For example, one can buy table top sources which deliver more than 100 TW (2.5 J in 25 fs). A laser beam of this kind focused on a 10$^{-6}$ m$^2$ spot gives an intensity of 10$^{20}$ W.m$^{-2}$ which corresponds to an energy density $\epsilon_0 E_\omega^2 + B_\omega^2/\mu_0$ such that the electric field $E_\omega$ is about 1.36 $\times$ 10$^{11}$ V/m and the magnetic field $B_\omega$ is about 4.5  $\times$ 10$^2$ T. These values are of the order of the highest static fields ever obtained in laboratories, and therefore intense lasers can also be used as sources of electromagnetic fields to induce nonlinear optics effects. We label this kind of effects with the {\it optical field induced} prefix in their name.

\subsection{Electrostatic fields}

In principle, the external fields needed for experiments can be either magnetic or electric. In vacuum the same level of effect is obtained in the presence of a $B$ field or an electric field $E$ equal to $cB$. This means that to equalize an effect created by a magnetic field of 1 T, one has to use an electric field of 300 MV/m. It looks like that, from the technological point of view, magnetic fields of several Tesla are easier to produce than electric fields of about 1 GV/m. Experimentalists have therefore mostly concentrated their efforts to magnetically induced effects in vacuum.

To produce a magnetic field, the standard method is to let a current circulate in a coil. The obtained magnetic field is proportional to the current density. This current density creates also a force density which is proportional to the product of the current density and the magnetic field and hence to the square of the magnetic field. Moreover, the current creates losses in the conductors by Joule effect and it heats the system. These two effects have to be taken into account to reach high magnetic fields.

Several solutions exist. One is to avoid heating thanks to superconducting wires. This is the solution chosen at CERN for example for the Large Hadron Collider magnets \cite{LHCmag}. The maximum field achievable depends on the critical fields of the materials above which the superconductivity property disappears. Magnets around 20 T are commercially available.

Another solution is to remove the dissipate electrical energy. It can be done with water. By this way, one can achieve magnetic field up to 35 T  but one needs a big electrical installation of several tens mega watts. Only a few installations around the world have installed this electrical power : NHMFL-DC (Tallahassee, USA) \cite{NHMFL}, LNCMI-G (Grenoble, France) \cite{LNCMIG}, HFML (Niemegen, Neetherland) \cite{HFML}, HMFL (Hefei, China) \cite{HMFLHef}, TML, (Tsukuba, Japan) \cite{TML}. A third way is to produce magnetic field with a combination of the two last methods : superconducting and copper wires cooled with water. The actual record of this kind of coils is 45 T during one hour.

Another way to avoid the problem of heating is to use pulsed fields. The idea is to discharge a lot of energy during a small duration in a coil. In this way, one can reach 80 T in 10-100ms depending on the available bank of capacitors which delivers the energy. Up to date, the world record has been reached at the NHMFL-PF in Los Alamos in the United States with 100 T. The main worldwide pulsed field installations are NHMFL-PF (Los Alamos, USA) \cite{NHMFL-PF}, LNCMI-T (Toulouse, France) \cite{LNCMIT}, HLD (Dresden, Germany) \cite{HDL}, HFML (Niemegen, Neetherlands) \cite{HFML}, WHMFC (Wuhan, China) \cite{WHMFC}, TML (Tsukuba, Japan) \cite{TML}.
The limitation of these pulsed coils is a combination of heating and magnetic pressure which can reach intensities of 10$^9$ Pascal.

To go further and to reach 300 T in a few microseconds, a possibility is given by a single turn coil. This is called megaGauss installation and only 2 are operational around the world (Toulouse and Tokyo) (see e.g. \cite{MegaGauss}). In this experiment, a single turn coil is placed at the end of a generator which delivers more than 60 000 A in one microsecond. After a few microseconds the coil is destroyed but this duration is enough to perform optical measurements.

More recently, the LNCMI Toulouse has developed special coils and generators for bringing high magnetic fields to special external installations like particle accelerator facilities or intense laser facilities. The generators are transportable and then high magnetic fields can move almost everywhere (see e.g. \cite{TrGen}).

The highest fields of which we have discussed previously are longitudinal magnetic fields. This means that the magnetic field is parallel to the optical access like in the case of simple solenoids \cite{Jackson} : this is called Faraday configuration in reference to the Faraday effect which needs a longitudinal magnetic field with respect to the light propagation.

For some experiments, one needs a magnetic field transverse with respect to the optical access like in the case of Helmoltz coils \cite{Crosser2010}. This configuration is less conventional. The major difficulty lies on the possibility of putting efficient reinforcements against magnetic pressure because of the non cylindrical symmetry of this kind of coils. Superconducting dipole magnets at CERN are able to give 10 T over 10 m. At LNCMI-Toulouse, the laboratory has developed a special coil \cite{LNCMIT} dedicated to the observation of vacuum magnetic birefringence which has already produced more than 30 T over a length of 50 cm. This equipment can be used almost everywhere thanks to the transportable generators, and transverse coils designed to deliver up to 40\ T have also been developed at LNCMI-T for plasma experiments where a strong pulsed field is coupled to an intense pulsed laser at LULI (France) \cite{LULI}.

\section{Phenomenology and related experiments}

A very large number of phenomena are expected in a nonlinear optical medium. The nonlinear response can give rise to exchanges of energy between electromagnetic fields of different frequencies. Few of them are allowed in a quantum vacuum because of C,P,T and Lorentz invariances. The absence of linear terms in $E$ or $B$ in the energy density given in eq. (\ref{Ude}) means that vacuum cannot have a permanent electric or magnetic dipole moment which seems obvious. The quadratic terms in $E$ or $B$ in eq. (\ref{Ude}) corresponds to the Maxwell energy density. Any ulterior term of this kind can be canceled out by a renormalization of the velocity of light.

As already told in the introduction, we focus our attention to low energy effects that affect the propagation of light in a vacuum. We also restrict mainly to effects induced by fields that are small compared with the critical ones.
Generally speaking, we treat phenomena mostly in the approximation

\begin{equation}\label{appr}
   {\hbar \omega \over m_e c^2}{B \over B_{cr}} = {\hbar \omega \over m_e c^2}{E \over E_{cr}} << 1.
\end{equation}

In this approximation, one can restrict to the first terms of the development of the Heisenberg-Euler lagrangian. This lagrangian has not yet been tested experimentally and therefore any experiment which goal is the measurement of one of the following effects tests a pure QED fundamental prediction.

\subsection{Three-wave mixing}

\begin{figure}[h]
\begin{center}
  \includegraphics[width=8cm]{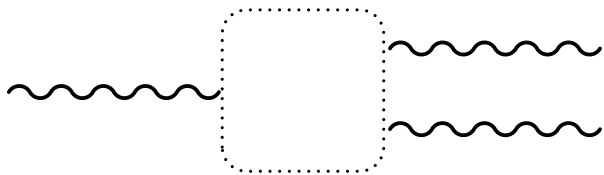}
\label{3waves}
    \end{center}
      \caption{Three-wave mixing describes an interaction between three electromagnetic waves. These effects are not allowed in vacuum.}
\end{figure}

Three-wave mixing indicates any term in eq. (\ref{Ude}) proportional to a product of three electromagnetic fields like $E^3$, $E^2B$, $EB^2$ and $B^3$.  No term containing three electromagnetic fields exists in eq. (\ref{Ude}), and therefore none of these effects is allowed in a vacuum.

In a nonlinear medium, these terms are linked to the second order nonlinear susceptibility and they include the optical rectification, the Faraday effect, the Pockels effect, the second harmonic generation or the parametric amplification \cite{Shen}.

A general three-wave mixing can be viewed as the generation of an optical wave by the combination of two different ones and viceversa (see figure 4).
Let the frequencies and wavevectors of these two optical waves be $(\omega_1,\bf{k_1})$, $(\omega_2,\bf{k_2})$, then the frequency and the wavevector of the third optical wave can be written as

\begin{equation}\label{3wa}
\omega_3=\omega_1\pm \omega_2\\
\bf{k_3}=\bf{k_1} \pm \bf{k_2}
\end{equation}

The optical rectification is the generation of a DC polarization or a DC magnetization in a nonlinear medium at the passage of an intense optical beam \cite{Bass1962} ($\omega_1=\omega_2$, $\omega_3=\omega_1-\omega_2=0$). It is a special case of difference frequency generation \cite{Shen} since it can be interpreted as initial photons forming new photons of zero energy and frequency.

The second harmonic generation, also called frequency doubling, is a process in which photons interacting with a nonlinear material to form new photons of twice the energy, and therefore twice the frequency of the initial photons \cite{Franken1961} ($\omega_3=2\omega_1$). It is a special case of sum frequency generation \cite{Shen}.

Optical Parametric amplification involves the transfer of power from a "pump" wave at $\omega_3$ to two waves at lower frequencies $\omega_1$ and $\omega_2$, with $\omega_3=\omega_1+\omega_2$ \cite{Yariv}. In particular, a photon interacting with a nonlinear material may give rise to two photons each of half the energy of the incoming one \cite{Giordmaine}.

The Faraday effect is the rotation of the plane of polarization of a linearly polarized light beam which is linearly proportional to the component of the magnetic field in the direction of propagation \cite{Faraday1845}. Faraday effect can be associated to a circular birefringence i.e. to a difference of the index of refraction for light rightward or leftward circularly polarized with respect to the direction of the magnetic field \cite{Shen}.

The Pockels effect, also known as the electro-optic effect, is the linear birefringence in an optical medium induced by an electric field \cite{Pockels1893}. Linear birefringence means that the index of refraction depends on the linear polarization of light \cite{Shen}.

As said before, none of these effects exists in vacuum.

\subsection{Four-wave mixing}

\begin{figure}[h]
\begin{center}
  \includegraphics[width=8cm]{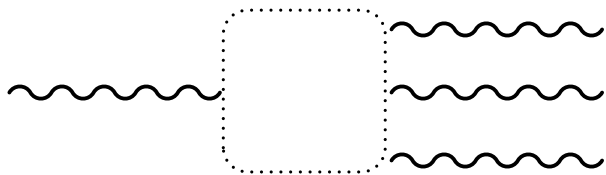}
  \caption{Four-wave mixing. It represents the combination of four electromagnetic waves. These effects are allowed in vacuum.}\label{4waves}
    \end{center}
\end{figure}

Four-wave mixing is a general name representing any effect due to the combination of four electromagnetic fields which can be collinear or not. Some of them can be of zero frequency i.e. electrostatic fields. It corresponds to the terms proportional to $E^4$, $E^2B^2$, and $B^4$ in eq. (\ref{Ude}). As shown in fig. \ref{4waves}, this can be view as a combination of two waves to give two different ones, or a combination of three of them to give one and viceversa.
Four-wave mixing is allowed in quantum vacuum.

\subsubsection{Vacuum nonlinear static polarization or magnetization\\}

Let's first of all present the case where only static fields are involved (see fig. 6).  Eq. (\ref{Pol}) and Eq. (\ref{Magn}) become:

\begin{equation}\label{PolStat}
\textbf{P}_0 = 4c_{2,0}
\epsilon_{0}\textbf{E}_0(\epsilon_{0}E^2_0 - {B^2_0 \over \mu_{0}}) + 2c_{0,2} {\epsilon_{0}
\over \mu_{0}} \textbf{B}_0(\textbf{E}_0 \cdot \textbf{B}_0),
\end{equation}

\begin{equation}\label{MagStat}
\textbf{M}_0 = -4c_{2,0}
{\textbf{B}_0 \over \mu_0}(\epsilon_{0}E^2_0 - {B^2_0 \over \mu_{0}}) + 2c_{0,2} {\epsilon_{0}
\over \mu_{0}} \textbf{E}_0(\textbf{E}_0 \cdot \textbf{B}_0),
\end{equation}

where $E_0$ and $B_0$ are the static electric and magnetic field, respectively. These formulas clearly indicate that a vacuum is polarized and magnetized by the presence of static fields. It is important to stress that $P_0$ not only depends on $E_0$ but also on $B_0$, and $M_0$ not only depends on $B_0$ but also on $E_0$. More over $P_0$ and $M_0$ depend as well on the angle between $E_0$ and $B_0$.

For the sake of argument, let's estimate the value of magnetization expected when only  $B_0$ is present. Eq. (\ref{MagStat}) gives $\mu_{0}{M}_0 \approx 5.3 \times 10^{-24}\times B_0(\textrm{T})^3$, a value that looks out of reach since nowadays the best magnetometers are not able to measure less than $10^{-15}$ T in one second (see e.g. \cite{Drung2007}).

\begin{figure}[h]
\begin{center}
  \includegraphics[width=4cm]{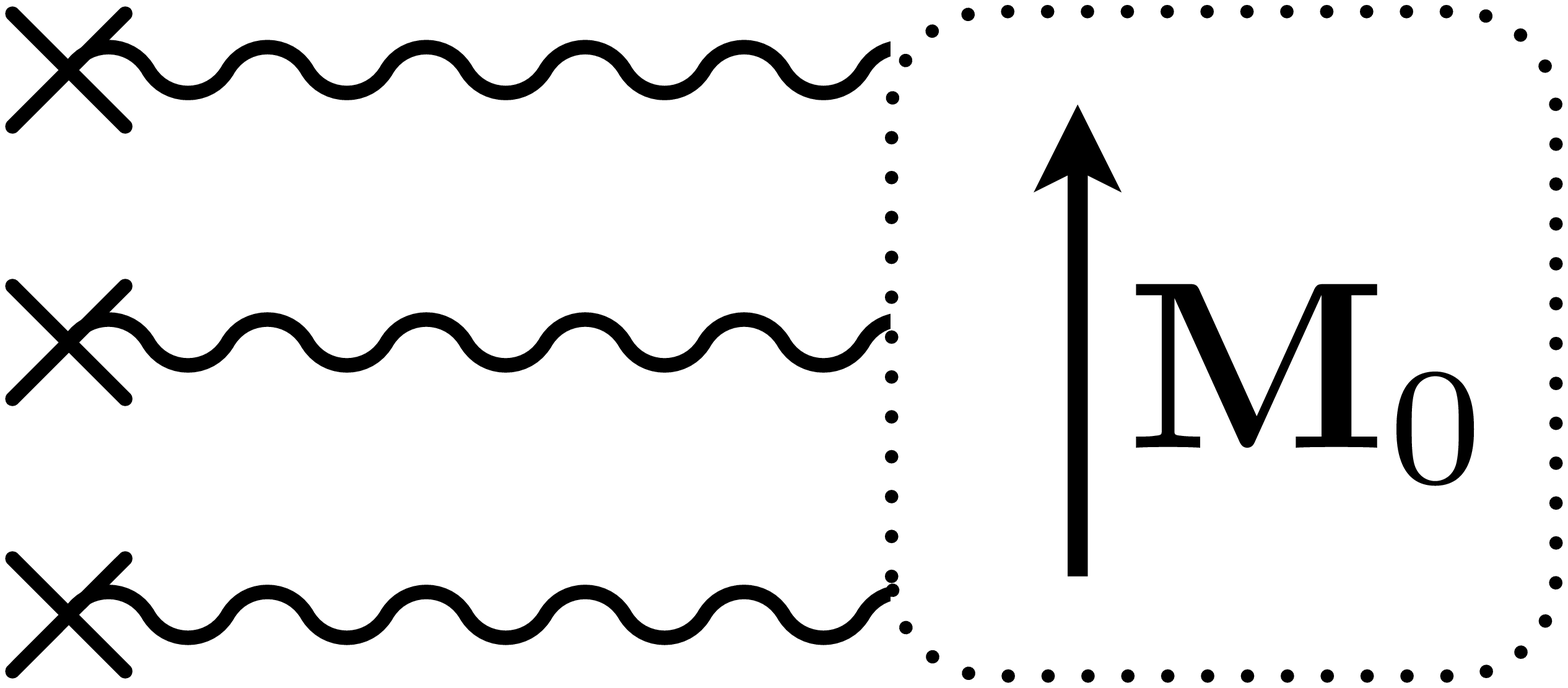}
  \caption{Four-wave mixing: three interactions of an electrostatic field give rise to a static magnetization.}\label{M4waves}
    \end{center}
\end{figure}

In ref. \cite{Kabat2002} and ref. \cite{Cohen2009} the expected magnetization in the presence of a external magnetic field is given in the framework of QCD. This value of magnetization for a magnetic field $B_0 \ll B^{QCD}_{cr}$ scales as the forth power of ratio between the electron mass and the pion mass which is about 280. A magnetization $\mu_{0}{M}_0 \approx 10^{-33}\times B_0(\textrm{T})^3$ is predicted \cite{Kabat2002}.

\subsubsection{Kerr effect, Cotton-Mouton effect, Jones birefringence, Magneto-electric birefringence\\}

In this section we deal with linear birefringences in vacuum. These birefringences can be induced by an electric field, a magnetic field or a combination of both. All these birefringences are manifestations of four wave mixing when two of the waves are static fields.

Following eq. (\ref{Pol}), we can write :

 \begin{eqnarray}\label{PolBir}
\textbf{P} &=& 4c_{2,0} \epsilon_{0}(\textbf{E}_{\omega}+\textbf{E}_0)\left(\epsilon_{0}E^2_0 - {B^2_0 \over \mu_{0}}+2\epsilon_0\textbf{E}_{\omega} \cdot \textbf{E}_0-{2 \textbf{B}_{\omega} \cdot \textbf{B}_0 \over \mu_0}\right)
 +\\
\nonumber  & & 2c_{0,2} {\epsilon_{0} \over \mu_{0}} (\textbf{B}_\omega+\textbf{B}_0)(\textbf{E}_\omega \cdot \textbf{B}_0+\textbf{E}_0 \cdot \textbf{B}_\omega+\textbf{E}_0 \cdot \textbf{B}_0),
\end{eqnarray}

\begin{eqnarray}\label{MagBir}
\textbf{M} &=& -4c_{2,0} {(\textbf{B}_{\omega}+\textbf{B}_0) \over \mu_0}\left(\epsilon_{0}E^2_0 - {B^2_0 \over \mu_{0}}+2\epsilon_0\textbf{E}_{\omega} \cdot \textbf{E}_0-{2 \textbf{B}_{\omega} \cdot \textbf{B}_0 \over \mu_0}\right)
 +\\
\nonumber  & & 2c_{0,2} {\epsilon_{0} \over \mu_{0}} (\textbf{E}_\omega+\textbf{E}_0)(\textbf{E}_\omega \cdot \textbf{B}_0+\textbf{E}_0 \cdot \textbf{B}_\omega+\textbf{E}_0 \cdot \textbf{B}_0),
\end{eqnarray}

where we have written the total electromagnetic fields $\textbf{E}$ and $\textbf{B}$ as the sum of the fields associated to the propagating waves $\textbf{E}_\omega$ and $\textbf{B}_\omega$ and the static ones, assuming that $E_\omega << E_0$ and $B_\omega << B_0$. We assume that the static fields are also homogeneous. A recent study of photon propagation in vacuum in non homogeneous fields is reported in ref. \cite{Karbstein2012}.

\begin{figure}[h]
\begin{center}
  \includegraphics[width=8cm]{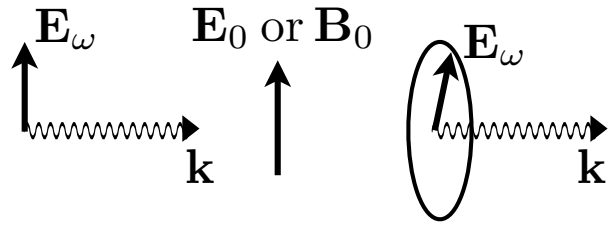}
  \caption{Kerr effect or Cotton Mouton effect. A linear birefringence is induced by a static field (electric or magnetic) perpendicular to the direction of light propagation. A linear polarization is converted into an elliptical polarization thanks to the presence of a static field.}\label{KE}
    \end{center}
\end{figure}

Since 1875 it is known that a static transverse electric field can induce a linear birefringence in a medium. This effect is called Kerr effect from the name of the physicist who discovered it \cite{Kerr1875}. The index of refraction depends on the light linear polarization with respect to the direction of the applied electric field. The difference $\Delta n_K$ between the index of refraction for light polarized parallel to the electric field $n_\|$ and the index of refraction for light polarized perpendicular to the electric field $n_\bot$ is proportional to $E_0^2$, $\Delta n_K = k_{K} E_0^2$.

At the turn of the century it was experimentally shown that, when linearly polarized light propagates in the
presence of a magnetic field $B_0$ normal to the direction of
light, media show also a birefringence similar to the Kerr one and one can write that $\Delta n_{CM} = k_{CM} B_0^2$. This magnetic field induced
linear birefringence is usually called Cotton-Mouton effect (CME)
since it was first investigated in detail by A.Cotton
and H.Mouton \cite{Cotton} since 1905 (see fig. \ref{KE}).

\begin{figure}[h]
\begin{center}
  \includegraphics[width=8cm]{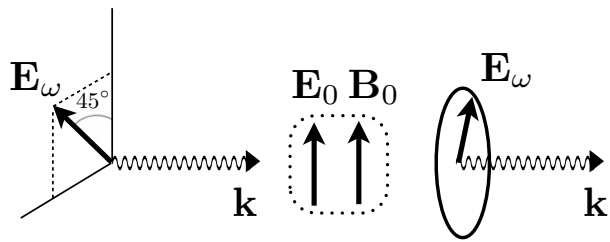}
  \caption{Jones linear birefringence. A linear birefringence is induced by both electric and magnetic fields perpendicular to the direction of light propagation. The birefringence axis are at $\pm$ 45$^\circ$ with respect to the static fields.}\label{JLB}
    \end{center}
\end{figure}

\begin{figure}[h]
\begin{center}
  \includegraphics[width=8cm]{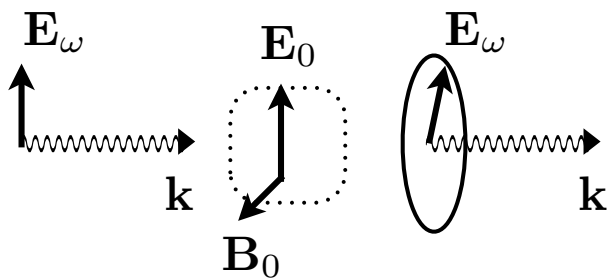}
  \caption{Magneto-electric linear birefringence. A linear birefringence is induced by crossed electric and magnetic fields, both perpendicular to the direction of light propagation. If one of the static fields is parallel to the direction of light propagation, the magneto-electric birefringence vanishes. One gets only a birefringence due to the field perpendicular to the direction of light propagation.}\label{MLB}
    \end{center}
\end{figure}

More recently the existence of a linear birefringence induced by the combined effect of a transverse electric field $E_0$ and a transverse magnetic field $B_0$ has been proven, both when $\textbf{E}_0 \| \textbf{B}_0$ (Jones linear birefringence) \cite{Roth2000} (fig. \ref{JLB}) and when $\textbf{E}_0 \bot \textbf{B}_0$ (Magneto-electric linear birefringence)\cite{Roth2002} (fig. \ref{MLB}). For this last kind of birefringence the birefringence axis are given by the electric and magnetic static fields.  Jones birefringence has the particularity that the birefringence axis  are at $\pm45^\circ$ with respect to the static fields instead of parallel and perpendicular as for Kerr and Cotton-Mouton effect. For Jones birefringence the difference $\Delta n_J$ between the index of refraction for light polarized at $+45^\circ$ to the electric field $n_+$ and the index of refraction for light polarized at $-45^\circ$  to the electric field $n_-$ is proportional to $E_0B_0$, $\Delta n_J = k_{J} E_0B_0$. For magneto-electric birefringence one can write that $\Delta n_{ME} = n_B-n_E = k_{ME} E_0B_0$. Symmetry considerations indicate that $k_{J}=k_{ME}$ \cite{Ross1989}. For both bilinear birefringences, another particularity is that $n$ changes sign when the wavevector of light $\textbf{k}$ becomes $-\textbf{k}$. A medium immersed in a magnetic and electric field therefore also show an axial birefringence since $\Delta n_a = n_{+k}-n_{-k} = k_{a} E_0B_0$ as proved experimentally for the first time in ref. \cite{Rikken2002}.

All these phenomena are expected in a vacuum. Therefore, in the presence of
electrostatic fields quantum vacuum behaves as a uniaxial birefringent crystal.
However, as it is shown in ref. \cite{Rizzo2005}, in the presence of both electric and magnetic fields perpendicular to each other with one of these fields parallel to the direction of light propagation, no bilinear birefringence appears, but only a birefringence due to the Kerr or the Cotton Mouton effect.

The reference papers for the value of the magnetic Cotton-Mouton birefringence and of the electric Kerr birefringence are ref. \cite{Bialynicka-Birula1970} and ref. \cite{Adler1971}. More recently, in 2000, the effects due to the presence of an electric field $E_0$ and a magnetic field $B_0$ have been studied in ref. \cite{Denisov2000} and in ref. \cite{Rikken2000} where the connection with the magnetoelectric and Jones birefringence is presented.

Following \cite{Bakalov1998}, let's study the Cotton-Mouton effect which is the vacuum nonlinear optics effect that has mostly attracted the interest of experimentalists.  This effect is due to the presence of a static transverse magnetic field \textbf{$B_0$}. Neglecting static terms, the equations (\ref{PolBir}) and (\ref{MagBir}) become :

\begin{equation}
\textbf{P}^{CM} =-4c_{2,0} \epsilon_{0} {B^2_0 \over \mu_{0}} \textbf{E}_{\omega} + 2c_{0,2} {\epsilon_{0} \over \mu_{0}} \textbf{B}_0(\textbf{E}_\omega \cdot \textbf{B}_0),
\end{equation}

\begin{equation}
\textbf{M}^{CM} = 4c_{2,0} {B^2_0 \over \mu_{0}^2}\textbf{B}_\omega
 + 8c_{2,0}  {\textbf{B}_0 \over \mu_{0}^2}(\textbf{B}_\omega \cdot \textbf{B}_0)
\end{equation}

and one can deduce that

\begin{equation}
\epsilon_{\|} =  \epsilon_{0} - 4{c_{2,0}}{\epsilon_{0} \over
\mu_{0}} B_{0}^2 + {2c_{0,2}}{\epsilon_{0} \over \mu_{0}}
B_0^2,
\end{equation}

\begin{equation}
\epsilon_{\bot} =  \epsilon_{0} - 4{c_{2,0}}{\epsilon_{0} \over
\mu_{0}} B_{0}^2,
\end{equation}

\begin{equation}
\Delta \epsilon = \epsilon_{\|} - \epsilon_{\bot} =
 {2c_{0,2}}{\epsilon_{0} \over \mu_{0}}
B_0^2
\end{equation}

and in the same way one also has

\begin{equation}
\mu_{\|} =  \mu_{0} (1 + 4{c_{2,0}}{1 \over
\mu_{0}^2} B_{0}^2),
\end{equation}

\begin{equation}
\mu_{\bot} =  \mu_{0} (1 + 12{c_{2,0}}{1 \over
\mu_{0}^2} B_{0}^2),
\end{equation}

\begin{equation}
\Delta \mu = \mu_{\|} - \mu_{\bot} = - 8{c_{2,0}} B_{0}^2,
\end{equation}

where the symbol $\|$ accompanies any
quantity related to a light polarization parallel to the static
field, and the symbol $\bot$ any quantity
related to a light polarization perpendicular to the static
field.

Then one can calculate the refractive index of the medium :

\begin{equation}
n_{\|} = {\sqrt{\epsilon_{\|} \mu_{\|}} \over \sqrt{\epsilon_{0}
\mu_{0}}} =  1 + {c_{0,2}}{B_{0}^2 \over \mu_{0}},
\end{equation}

and

\begin{equation}
n_{\bot} = {\sqrt{\epsilon_{\bot} \mu_{\bot}} \over
\sqrt{\epsilon_{0} \mu_{0}}} =  1 + {4c_{2,0}}{B_{0}^2 \over \mu_{0}},
\end{equation}

Finally the anisotropy $\Delta n$ is equal to

\begin{equation}
\Delta n_{CM} = n_{\|} - n_{\bot} = {\sqrt{\epsilon_{\|} \mu_{\|}} -
\sqrt{\epsilon_{\bot} \mu_{\bot}} \over \sqrt{\epsilon_{0}
\mu_{0}}} =  (c_{0,2} -4 c_{2,0}){B_0^2
\over \mu_{0}}
\end{equation}

Let's note that $n_{\|}$ depends only on $c_{0,2}$ and $n_{\bot}$
on $c_{2,0}$ like $\Delta \epsilon$ and $\Delta \mu$, respectively.
Let's also note that, since the velocity of light has to be anyway
smaller than $c$, $c_{0,2}$ and $c_{2,0}$ have to be positive.

The result given in the previous equation holds as far as Lorentz invariance holds. QED prediction via the Heisenberg-Euler lagrangian is that $c_{0,2} = 7 c_{2,0}$ and therefore one can finally write that

\begin{equation}
\Delta n_{CM} = 3c_{2,0}{B_0^2\over \mu_{0}}.
\end{equation}

Taking also into account Ritus corrections for $c_{0,2}$ and $c_{2,0}$, one obtains a more precise result:

\begin{equation}
\Delta n_{CM} = \left({2\alpha^2 \hbar^3 \over 15 m_{e}^4 c^5} + \frac{5}{6}\frac{\alpha^3 \hbar^3}{\pi m_{e}^4 c^5}\right) {B_0^2 \over \mu_{0}}={2\alpha^2 \hbar^3 \over 15 m_{e}^4 c^5}\left(1+ \frac{25\alpha}{4\pi}\right){B_0^2 \over \mu_{0}}.
\end{equation}

Ritus term for $\Delta n_{CM}$ corresponds to a $1.45 \%$ correction to the leading term.

Finally, using CODATA values \cite{CODATA} for fundamental constants one obtains
$k_{CM} = (4.0317\pm0.0009)\times10^{-24}$ T$^{-2}$
where the uncertainty is calculated assuming arbitrarily that the $\alpha^2$ order radiative correction, which has never been calculated, amounts to about a $1.5\%$ correction like the $\alpha$ order one.

In ref. \cite{Tsai1975} and more recently in ref. \cite{Heyl1997_3} the indexes of refraction $n_{\|}$ and $n_{\bot}$ are given when the static magnetic field $B_0$ is comparable and even greater of $B_{cr}$. It is worth stressing that linear birefringence effects can be considered achromatic as long as ${\hbar \omega \over m_e c^2}{B \over B_{cr}} = {\hbar \omega \over m_e c^2}{E \over E_{cr}} << 1$.

In ref. \cite{CasteloFerreira2008} one can also find corrections to the value of $\Delta n_{CM}$ calculated in the framework  of QCD. As already discussed these corrections are negligeable because of the QCD energy scale.

The values of the Cotton-Mouton, Kerr and magnetoelectric birefringences are all related by vacuum Lorentz invariance
as demonstrated in ref. \cite{Rizzo2005}. Kerr effect must have the same value for the $\Delta n_{K}$ than the Cotton-Mouton effect but opposite sign for $E_0=c B_0$.

\begin{equation}
\Delta n_K = -{2\alpha^2 \hbar^3 \over 15 m_{e}^4 c^5}\left(1+ \frac{25\alpha}{4\pi}\right) \epsilon_0 E_{0}^2,
\end{equation}
i.e. $k_{K} \approx -4.4\times10^{-41}$ m$^2$.V$^{-2}$. The value of the magneto-electric birefringence must be twice the value of Cotton-Mouton birefringence when $E_0=c B_0$. The sign is negative if $\textbf{k} \| (\textbf{E}_0 \times \textbf{B}_0)$ .

\begin{equation}
\Delta n_{ME} =  -{4\alpha^2 \hbar^3 \over 15 m_{e}^4 c^5}\left(1+ \frac{25\alpha}{4\pi}\right) \sqrt{\epsilon_0 \over \mu_0} {\textbf{k} \over k} \cdot (\textbf{E}_0 \times \textbf{B}_0),
\end{equation}
i.e. $k_{ME} \approx \pm2.6\times10^{-32}$ m.V$^{-1}$.T$^{-1}$.
Finally Jones birefringence must have the same magnitude and sign as the magneto-electric one and therefore
$k_{J} \approx \pm2.6\times10^{-32}$ m.V$^{-1}$.T$^{-1}$ as well.

In both magneto-electric and Jones configurations, light going back and light going further in a vacuum do not travel at the same velocity since $n$ is different. This gives an axial birefringence $\Delta n_a$. One can show \cite{Rikken2003} that the axial birefringence is

\begin{equation}
\Delta n_{a} =n(\textbf k)-n(-\textbf k)=  -{8\alpha^2 \hbar^3 \over 15 m_{e}^4 c^5}\left(1+ \frac{25\alpha}{4\pi}\right) \sqrt{\epsilon_0 \over \mu_0} ({E}_0 {B}_0),
\end{equation}

i.e. $k_a =  5.2\times10^{-32}$ m.V$^{-1}$.T$^{-1}$.

Experiments looking for a variation of the velocity of light in the presence of a magnetic field
can be traced back to the end of the XIX century when Morley, Eddy and Miller somewhat inaugurate this
experimental field using a Michelson-Morley interferometer \cite{BornWolf} in 1898 \cite{Morley1898}.
\begin{figure}[h]
\begin{center}
  \includegraphics[width=10cm]{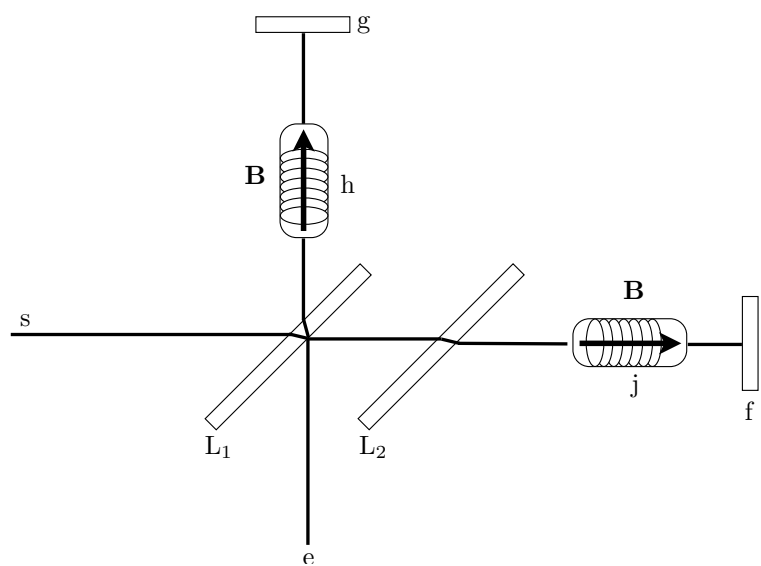}
  \end{center}
  \caption{Michelson and Morley experiment. A train of waves of light emitted from the source s is divided in two beams which are back reflected on mirrors $g$ and $f$. Both waves are recombined on L$_1$ and interferences bands can be seen by the eyes at $e$ if the distances of $f$ and $g$ from L$_1$ are properly adjusted. Then two tubes filled with carbon bisulphide $h$ and $j$ are put in the paths of the divided ray. Each tube is surrounded with a copper coil in order to create a longitudinal magnetic field.}
  \label{Morley}
\end{figure}

The basic idea (see fig. \ref{Morley}) was to split in two a beam of light coming from a flame of a Bunsen lamp colored by the introduction of some sodium compound, polarized by a Nicol prism\cite{BornWolf}, and to observe the interference band due to the different optical paths in the two interferometer arms.
If light velocity was affected by the presence of a 0.165 T magnetic field in one of the two arms, a displacement of the interference band would have been observed. Fringes displacement was simply monitored by eyes. As a matter of fact, light did not propagate in vacuum but in liquid carbon bisulphide and air and the magnetic field was directed parallel to the light propagation vector.  Authors were looking for a retardation effect which should exist together with the Faraday rotation induced by the material medium. Such an effect is difficult to explain in modern terms. Nevertheless, they reached a sensitivity of one part in a hundred million in the measurement of a possible change of the velocity of light depending on the magnetic field.

First experiment clearly devoted to detecting a change in the propagation of light in vacuum induced by a transverse magnetic field has been performed by Watson in 1929 \cite{Watson1929}. The motivation was to look for a magnetic moment of the photon. The experimental method was based on the hypothesis that if a photon had a magnetic moment $\mu$, its energy in the presence of a magnetic field $H$ would change by a quantity $\Delta E =\pm\mu H$ because of the coupling of the field and the magnetic moment parallel or antiparallel to the field itself. As a consequence, photon wavelength would change by a quantity $\Delta \lambda = \pm\mu H\lambda^2/hc$. The experimental arrangement (see fig. 11) consisted of a neon discharge tube as light source, a Nicol prism\cite{BornWolf} as a polarizer, and a Fabry-Perot interferometer\cite{BornWolf} to detect the change in $\lambda$. This change would appear as a change of the interference pattern. The magnetic field intensity was  1 T over 0.02 m, and it was directed perpendicularly to the light direction of propagation. Interference pattern with and without the magnetic field was captured by a photographic camera. The region inside the 0.01 m long Fabry-Perot interferometer was evacuated. The author stated that the alteration in the refractive index of vacuum produced by a magnetic field perpendicular to the direction of light propagation does not exceed 4 $\times$ $10^{-7}$ per Tesla. Nowadays, we know that because of the Cotton-Mouton effect of Vacuum the expected variation of light velocity is quadratic in the magnetic field and is of the order of $10^{-24}$.

Motivated again by the search for a photon magnetic moment and by Watson's paper \cite{Watson1929}, Farr and Banwell in 1932 \cite{Farr1932} and 1940 \cite{Banwell1940} reported measurements of the velocity of propagation of light in vacuum in a transverse magnetic field. Their experiment was first based on a Jamin interferometer\cite{BornWolf} and in the 1940 version on a Michelson interferometer \cite{BornWolf}.

\begin{figure}[h]
\begin{center}
  \includegraphics[width=10cm]{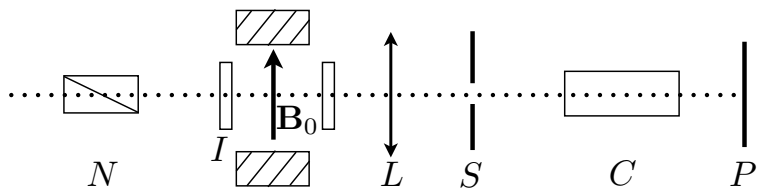}
  \end{center}
  \caption{Watson's experiment. Light coming from a neon discharge tube ($N$) passes through a Nichols prism $N$ and a transverse magnetic field produced by an electromagnet. A Fabry Perot interferometer $I$ is placed on the beam path and a lens $L$ image the fringes pattern on a photographic camera P through slits $S$ and collimator $C$.}
  \label{Morleybis}
\end{figure}

Light from an incandescence lamp was polarized by a Nicol prism \cite{BornWolf}, separated in two and then one of the rays passed through a magnetic field of 1.8 T over  1.125 m. Observation of a possible fringe displacement induced by the magnetic field was first observed by eyes. In the 1940 version of their apparatus the detector was a photoelectric cell, the signal of which was amplified and read by a galvanometer to record it on a moving photographic strip which increased the apparatus sensitivity. The final result was that in a 2 T field the relative variation of light velocity was less than  2 $\times$ $10^{-9}$ \cite{Banwell1940}.

No clear theoretical predictions were motivating such experiments. First citation of the existence in a vacuum of the
Cotton-Mouton effect can only be found in a paper by Erber in 1961 \cite{Erber1961} which also provides an estimation
of the value of the Cotton-Mouton $\Delta n_{CM}$, and where a discussion of different experimental approaches can also be found. This renewed interest was motivated by the laser invention \cite{Maiman} and the progresses in magnets providing fields of several Tesla. This has raised the hope that effects due to vacuum polarization in the presence of a strong magnetic field could be soon observable.

The idea of using a Michelson-Morley interferometer to measure the variation of velocity of light induced by a magnetic field has been proposed again several times \cite{GrassiStrini1979}, \cite{Ni1991}, \cite{Zavattini2009}, \cite{Dobrich2009} following the technical progresses in Michelson interferometry driven by the search for gravitational waves on earth \cite{LIGO},\cite{VIRGO} using such an apparatus.

The theoretical results published in the seventies \cite{Bialynicka-Birula1970}, \cite{Adler1971} raised a new interest in the field and in 1979 Iacopini and Zavattini proposed to measure the vacuum linear magnetic birefringence via the ellipticity
induced on a linearly polarized laser beam by the presence of a
transverse magnetic field \cite{Iacopini1979}. Actually, a linearly polarized light beam propagating in a birefringent medium becomes elliptically polarized. In general, the acquired ellipticity $\Psi$ is related to a birefringence $\Delta n$ by the formula:

\begin{equation}\label{ell}
\Psi = \pi \frac{L}{\lambda} \Delta n \sin 2\theta ,
\end{equation}

where $L$ is the optical path in the birefringent medium, $\lambda$ is the light wavelength and $\theta$ the angle between the light polarization and the birefringence axis \cite{BornWolf}. If $L$ is considerably greater than $\lambda$, the measurement of $\Psi$ can become an interesting indirect method to measure $\Delta n$. The measurement of the ellipticity is the main method to study Cotton-Mouton effect since its discovery in 1902 \cite{Rizzo1997}. Since $\Psi$ is proportional to $\Delta n$, in the case of vacuum Cotton-Mouton effect $\Psi \propto \left(\frac{B_0}{B_{cr}}\right)^2$.

It is worth mentioning that the propagation in a birefringent medium not only induces an ellipticity, but it also rotates the light polarization direction \cite{BornWolf}. If $\Psi \ll 1$ the rotation angle $\theta_r$ can be written as $\theta_r \leq \frac{\Psi^2}{2}$, therefore one can assume that $\theta_r \propto \left(\Delta n\right)^2$, and in particular for vacuum Cotton-Mouton effect $\theta_r \propto \left(\frac{B_0}{B_{cr}}\right)^4$. This means that this rotation can be safely neglected.

Zavattini's proposal has been a big step forward since all the experiments performed in recent years or under way are based on it. Let's discuss it in details.

\begin{figure}[h]
\begin{center}
  \includegraphics[width=10cm]{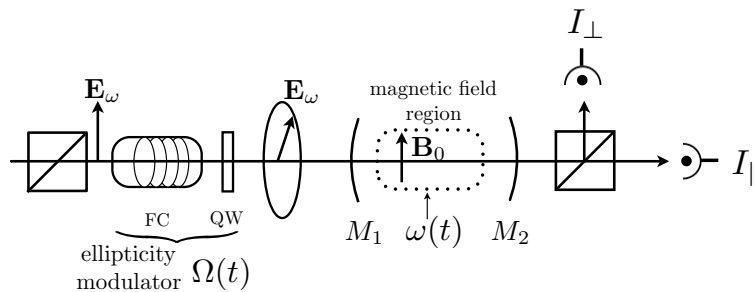}
  \end{center}
  \caption{Zavattini's proposal for ellipticity measurement. A linearly polarized beam is injected in a Fabry-Perot cavity formed by the two mirrors $M1$ and $M2$, in which a magnetic field $\vec B_0$ is applied. The induced birefringence (Cotton Mouton effect) transforms the linear polarization in an elliptical polarization. The induced ellipticity is enhanced by an optical cavity because the effect can be summed when light goes back and forth in the magnetic field. The apparatus is put between crossed polarizer and analyzer in order to measure the ellipticity given by the ration $I_{\perp}/I_{\parallel}$. Moreover one can modulate the field and use an ellipticity modulator for measuring the effect with an heterodyne detection.  }\label{zav}
\end{figure}

In order to have a larger optical path in the field the effect to be measured is increased using an optical cavity. Moreover the ellipticity and the magnetic field were modulated in order to be able to use heterodyne detection technique to increase the signal to noise ratio. As sketched in figure \ref{zav}, a linearly polarized beam passes through an ellipticity modulator composed by a Faraday cell and a quarter wave plate properly aligned. The magnetic field of the faraday cell gives a modulated ellipticity  $\eta(t)=\eta_0 \cos (\Omega t)$, which induces an ellipticity modulation. Then the beam is injected in a transverse magnetic field surrounded by an optical cavity defined by mirrors $M_1$ and $M_2$.  The magnetic field is also modulated so that the ellipticity induced by the field can be written as $\psi(t)=\psi_0 \cos(\omega t + \phi)$. At the output of the optical cavity, the light beam is analyzed with a polarizer ($A$) crossed with respect to the first one ($P$). The light is sent to two photodiodes which deliver signals proportional to $I_\parallel$ and $I_\perp$. The transmitted light after the optical cavity is :
\begin{equation}
I_{\perp}=I_0[\sigma^2 + (\Theta+\psi(t)+\eta(t))^2],
\label{Iext}
\end{equation}
where $I_0$ is the beam intensity before the analyzer, $\sigma^2$ is the extinction factor of the polarizers, $\Theta$
represents any static uncompensated ellipticity.

Developing Eq. (\ref{Iext}), one sees that the transmitted light is composed of different frequency components (see table \ref{Fourier}).

\begin{table}[htdp]
\begin{center}
\begin{tabular}{|c|c|c|}
\hline
Frequency &  Intensity/I$_0$ & \\
\hline
DC & $\sigma^2+\Theta^2+\eta_0^2/2+\Psi_0^2/2$&\\
$2\Omega_0$ & $\eta_0^2$&\\
$\Omega_0\pm\omega$ & $\eta_0\psi_0$ & signal to be measured\\
$\Omega_0$ &$2I_0\Theta\eta_0$&\\
\hline
\end{tabular}
\end{center}
\caption{Fourier components of the transmitted light of interest for the heterodyne detection technique.}
\label{Fourier}
\end{table}%

Basically, one gives up the idea to measure directly the variation of light velocity as using a Michelson interferometer and one restricts himself to the measurement of $\Delta n_{CM}$. The advantage is the expected noise reduction with respect to the Michelson interferometer apparatus where interfering light follows two different paths.

After tests at CERN \cite{CERN} in Switzerland
\cite{Iacopini1981}, an apparatus has been set up at the
Brookhaven National Laboratory \cite{BNL}, USA \cite{Cameron1993}.
It is based on a multipass cavity for enhancing the signal by a factor 250, and a magnetic field up to 4 T over 8.8 m modulated at about 30 mHz. In ref. \cite{Cameron1993}, authors report a sensitivity in ellipticity of $7.9\times 10^{-8}$ rad/$\sqrt{\textrm{Hz}}$. This sensitivity being insufficient to detect the QED effect, authors concentrated their effort on the search for the existence of particles beyond the standard model coupling with electromagnetic fields.

Actually, in 1986, Maiani, Petronzio, and Zavattini \cite{Maiani1986} showed that hypothetical low mass, neutral, spinless bosons, scalar or
pseudoscalar, that couple with two photons could induce an ellipticity signal in the Zavattini apparatus similar to the one predicted by QED. Moreover, an apparent rotation of the polarization vector of the light could be observed because of conversion of photons into real bosons resulting in a vacuum
magnetic dichroism which is absent in the framework of standard
QED. The measurements of ellipticity and dichroism, including
their signs, can in principle completely characterize the
hypothetical boson, its mass $m_a$, the inverse coupling constant
$M_{a}$, and the pseudoscalar or scalar nature of the particle.
Maiani, Petronzio, Zavattini paper was essentially motivated by
the search for axions. These are pseudoscalar,
neutral, spinless bosons introduced to solve what is called the
{\it strong CP problem} i.e. the fact that there is no experimentally known violation of the CP-symmetry in quantum chromodynamics even if there is no known reason for it to be conserved in QCD specifically. A discussion about non standard model physics
in external fields can be found in ref.\cite{Gies2008}.

No signal was observed and the final result of the Brookhaven National Laboratory experiment was that $k_{CM} \leq 2.2 \times 10^{-19}$ T$^{-2}$.

In 1991, a new attempt to measure the vacuum magnetic
birefringence has been started at the LNL \cite{LNL} in Legnaro, Italy, by
the PVLAS collaboration \cite{Bakalov1998}. This experiment is again
based on ref. \cite{Iacopini1979}. A vertical Fabry-Perot cavity is
used to increase the effect to be measured, while a
superconductive 5 T magnet rotates around its own axis to modulate
it. To the first order, this case can be calculated in the approximation of regarding the magnetic field as fixed at its instantaneous angular orientation, using the standard vacuum birefringence formulae for a static magnetic field \cite{Adler2007}. Results on vacuum magnetic birefringence published in 2008 \cite{Zavattini2008} indicates that the apparatus had a noise level of about $1.7 \times 10^{-20}$ T$^{-2}$ for $k_{CM}$.

In the meantime a new proposal has been put forward based at National Tsing Hua University \cite{NTHU}, Hsinchu, Taiwan, Republic of China : the Q\&A 
project.
This project started around 1996 \cite{Ni1996}. The experimental set up is
similar to the PVLAS one but the magnetic field is produced by permanent magnets, the 3.5 m long cavity is formed by two high reflectivity mirrors suspended with two X-pendulum suspensions mounted on two isolated tables. No birefringence effect has been yet detected and the achieved sensitivity in ellipticity is 10$^{-6}$ rad/$\sqrt{\textrm{Hz}}$ \cite{Mei2010}.

Another proposal based on the use of pulsed magnets as suggested in ref \cite{Rizzo} has been presented in ref. \cite{Battesti2008} : the BMV project.
This experiment is also based on the Zavattini proposal and it is mounted at the LNCMI-T  \cite{LNCMIT}, Toulouse, France. The Fabry Perot cavity is 2.2 m long and the cavity finesse is greater than 400 000. The novelty of this experiment is the use of pulsed magnets in order to reach higher fields. A specially designed magnet delivers more than 10 T in a first version, and 30 T  has been already reached for the next generation experiment. In ref. \cite{BMVPRA} authors show that a single magnetic pulse is sufficient to detect birefringence signals as low as $k_{CM} \approx 5\times 10^{-20}$ T$^{-2}$.

The use of long superconducting magnets developed for accelerator machines has been also proposed by a collaboration \cite{Lee1995} based at Fermilab \cite{Fermilab}, USA, and more recently by a group based at CERN \cite{CERN} : OSQAR \cite{Pugnat2006}.

Finally very recently a new version of the PVLAS experiment has been set up at the INFN \cite{Ferrara} in Ferrara (Italy). As the Q\&A experiment, it is based on the use of rotating permanent magnets. The value of the field is 2.3 T. A Fabry-Perot cavity with a finesse of about 250 000 is used to increase the optical path. This apparatus has given a limit on vacuum magnetic linear birefringence of $k_{CM} \leq 4.4\times 10^{-21}$ T$^{-2}$ \cite{Zavattini2012}.

Ellipticity formula given before (Eq. (\ref{ell})) shows that decreasing the light wavelength increase the ellipticity effect to be measured. Erber discussed this experimental possibility in his 1961 paper \cite{Erber1961}, and more recently in ref. \cite{Cantatore1991} a proposal to use gamma rays to measure the vacuum magnetic birefringence can be found. Gamma rays are produced by inverse Compton scattering on an electron beam of a storage ring by a polarized laser beam. In principle the photons obtained are also polarized. After passing through a magnetic field, gamma polarization is analyzed by using the fact that shower production depends on gamma polarization in crystals.

The basic idea hidden in the Watson's apparatus of 1929 was to transform a light velocity variation into a frequency variation. Frequency measurements are among the most precise measurements that can be performed nowadays (see e.g. \cite{Chen2012}). Resonance frequency of optical cavities depends on the optical length of the cavity itself. Any variation of the index of refraction, for example induced by a magnetic field, will induce a variation of the resonance frequency. This gives the signal to be measured by beating it with a frequency reference signal. For vacuum Cotton-Mouton effect this idea has been first envisaged by Erber \cite{Erber1961}. More recently, the use of a ring He-Ne laser, i.e. a laser based on a ring optical cavity, is suggested in ref. \cite{Stedman1997}, a linear Fabry-Perot cavity in ref. \cite{Rizzo1997}, and a feasibility study of a vacuum magnetic birefringence by frequency shift measurement can be found in ref. \cite{Hall2000}. All methods give beat frequencies to be measured of the order of 10 nHz corresponding to a relative frequency shift of about $10^{-22}$ which looks very challenging since nowadays the best relative frequency measurements are at the level of about $10^{-17}$ (see e.g. \cite{Chen2012}).

As far as Kerr effect is concerned, let's recall that experiments in the presence of a static electric field looks from the technological point of view more difficult since to equalize the effect of a 1 T magnetic field one needs a 300 MV/m electric field. This is certainly the reason why the experimental observation of the Kerr effect in vacuum has never been tried yet.

Recently magneto-electric birefringence effects have also attracted the attention of experimentalists. The particularity of this type of effects is that they do not accumulate in a linear cavity as Cotton-Mouton and Kerr effects since the index of refraction depends on the direction of propagation with respect to the plane containing E and B vectors. One therefore needs a ring laser \cite{Denisov2000} or a ring cavity \cite{Rikken2000} (see fig. 13). Very recently an experimental method to measure magneto-electric axial birefringence in dilute matter has been proposed \cite{Robilliard2010} and a measurement performed in gas-phase molecular nitrogen has shown a noise level corresponding to a  $k_a \simeq 10^{-23}$ m.V$^{-1}$.T$^{-1}$ \cite{Pelle2011}. A ring cavity is used \cite{Stedman1997}, the injected laser beam is split in two and one looks to the difference in the resonance frequency between the laser beam turning clockwise and anticlockwise in the cavity.

\begin{figure}[h]
\begin{center}
  \includegraphics[width=6cm]{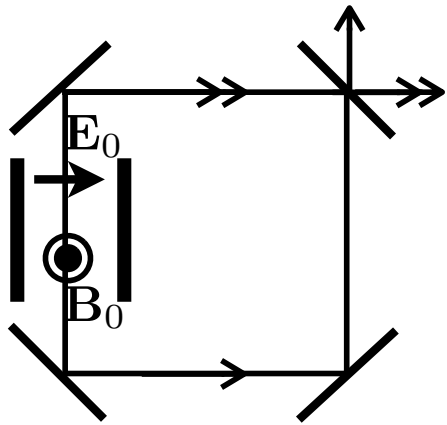}
  \end{center}
  \caption{Magneto-electric measurement. A ring Fabry Perot cavity is injected by two resonant beams. The induced magneto-electric birefringence depends on the direction of the light propagation with respect to the orientation of the electric and magnetic fields. Beams at the output of the interferometer are frequency shifted from each other because of magneto-electric birefringence. }\label{Denisov}
\end{figure}

\subsubsection{Optical field-induced birefringence\\}

In this kind of experiments the external field is produced by an electromagnetic wave.
One of the experimental challenges of this kind of measurements is to produce electromagnetic fields as high as possible and to have a long interaction region. Proposals have been published suggesting to use energetic laser pulses to this purpose (see fig. 14). As we have already discussed, in the focal spot of a powerful laser beam, electromagnetic fields may reach values exceeding the ones of static fields that can be found in laboratories.
\begin{figure}[h]
\begin{center}
  \includegraphics[width=8cm]{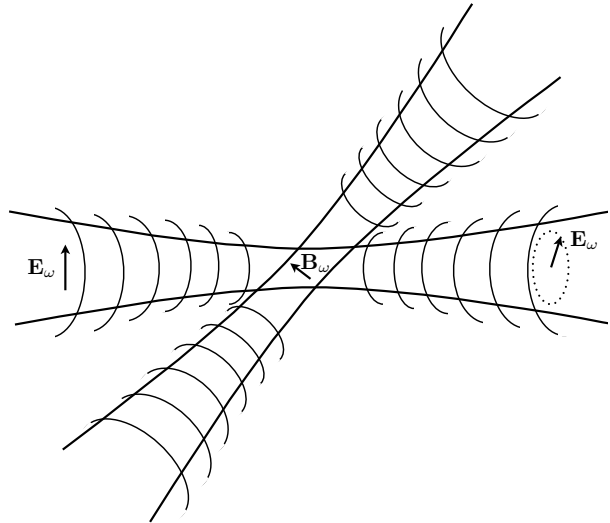}
  \end{center}
  \caption{Optical field-induced birefringence. The external electric or magnetic field is produced by an electromagnetic wave. In this kind of experiment, a linearly polarized beam becomes elliptically polarized passing through another electromagnetic wave.}\label{OFIB}
\end{figure}

A proposal to measure birefringence induced by a counter-propagating intense laser beam is reported in ref. \cite{Aleksandrov1985}.
In ref. \cite{Luiten2004} birefringence measurements using two pulsed laser beam, one as a probe and the second, more intense, as a field source is proposed and in ref. \cite{Homma2011}, in a similar configuration, the phase-contrast Fourier imaging \cite{PCFourier} is proposed as detection technique. In refs. \cite{Heinzl2006} and \cite{DiPiazza2006} a laser is again proposed as the field source while a x-ray beam is the probe to measure the birefringence taking advantage of the shorter wavelength. For example, authors of ref. \cite{DiPiazza2006} show that a wave of 0.4 nm wavelength propagating on a distance of 1.5 $\mu$m where a standing wave provide a laser intensity of 10$^{27}$ W.m$^{-2}$ acquires an ellipticity of about 4 $\times$ 10$^{-8}$ rad. It is important to note that because of diffraction effects, polarization direction is also rotated by an angle of the same order as the ellipticity. Further studies can also be found in ref. \cite{MarklundNP2010} and \cite{King2010NP}. Recently very high-purity polarization states of X-rays have been reported \cite{Marx2011}, which opens a possibility to detect the vacuum magnetic birefringence with this kind of experimental setup.

\subsubsection{Optical rectification induced by electrostatic fields i.e. inverse magneto-electric effects\\}

In a standard medium optical rectification can be induced by a light beam in the presence of an external magnetic and/or electric field. These effects are also known as inverse effects with respect to the field-induced birefringences.
\begin{figure}[h]
\begin{center}
  \includegraphics[width=5cm]{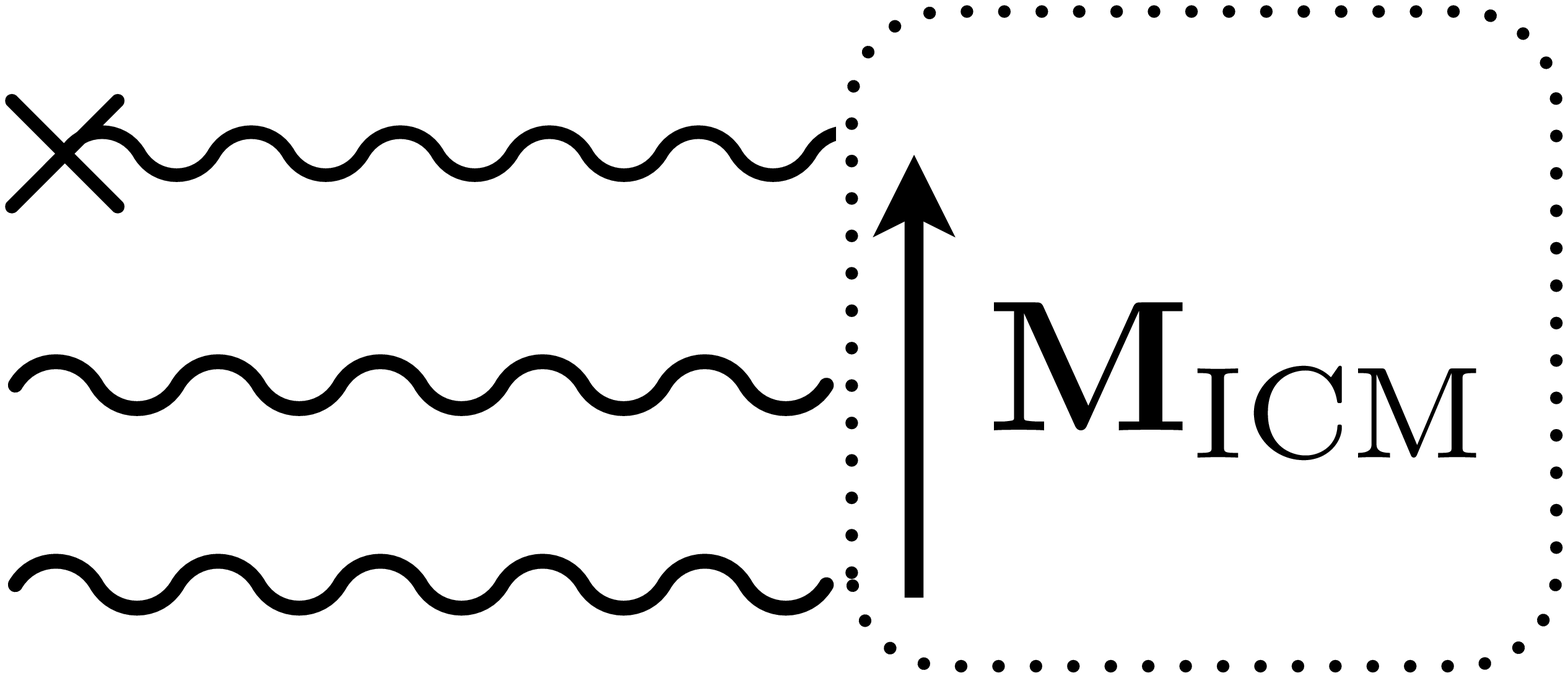}
 \end{center}
\caption{Inverse magneto electric effects. Optical rectification induced by the presence of an external magnetic or electric field. Here the interaction between two photons of the electromagnetic wave and a photon of the static magnetic field gives rise to a magnetization $\bf M$. This effect is known as Inverse Cotton Mouton effect if the static magnetic field is perpendicular with respect to the direction of light propagation.}
\end{figure}

For example, the Inverse Cotton-Mouton Effect (ICME) corresponds to a static magnetization induced in a medium by a non resonant linearly polarized light beam propagating in the presence of a transverse magnetic field (see fig. 15). This magnetization is proportional to the value of the magnetic field, and to the intensity of the propagating electromagnetic wave (see ref. \cite{Shen} and refs. therein).  As stated in ref. \cite{Shen}, microscopically, the light-induced dc magnetization arises in a standard medium because the optical field shifts the different magnetic states of the ground manifold differently, and mixes into these ground states different amount of excited states.
It looks like that inverse effects have not attracted much the attention of experimentalists. The observation of the ICME has only been reported very recently in a Terbium Gallium Garnet crystal  \cite{Baranga2011}. The ICME has also been calculated for the quantum vacuum in ref. \cite{Rizzo2010}.

Starting point of the calculation is eq. \ref{Magn}. Two cases are possible (${\bf E}_{\omega} \| {\bf B}_{0}$,
${\bf B}_{\omega} \bot {\bf B}_{0}$) or (${\bf E}_{\omega} \bot
{\bf B}_{0}$, ${\bf B}_{\omega} \| {\bf B}_{0}$).

In the first
case one gets:
\begin{equation}\label{mpar}
    {\bf M}_{\mathrm{ICM}\|} = 14c_{2,0} \epsilon_{0} E_{\omega}^2  {{\bf B}_{0} \over
\mu_{0}}= 14c_{2,0} {I_\omega \over c}  {{\bf B}_{0} \over
\mu_{0}}.
\end{equation}
In the second case one obtains:
\begin{equation}\label{mper}
    {\bf M}_{\mathrm{ICM}\bot} = 8c_{2,0} {B_{\omega}^2 \over \mu_{0}}  {{\bf B}_{0} \over
\mu_{0}} = 8c_{2,0} {I_\omega \over c} {{\bf B}_{0} \over
\mu_{0}}.
\end{equation}

where $I_\omega$ is the intensity associated to the electromagnetic wave, and where we have used the relation $c_{0,2}=7c_{2,0}$. In both cases ${\bf M}_{\mathrm{ICM}}$ is parallel to ${\bf B}_{0}$.

If a  laser pulse is focused to get $I_\omega \approx 10^{26}$\,W.m$^{-2}$ in a vacuum where a transverse magnetic field of more than 10 T is present,
the magnetization to be measured because of the ICME of the quantum
vacuum is

\begin{equation}\label{mparnum}
    {\bf M}_{\mathrm{ICM}\|} \approx 8 \times 10^{-11} \mathrm{T}
\end{equation}
and
\begin{equation}\label{mpernum}
    {\bf M}_{\mathrm{ICM}\bot} \approx 4.5 \times 10^{-11} \mathrm{T}
\end{equation}

where we have used the relation $\mu_0 \bf M$(A/m)= $\bf M$(T).
Authors of ref. \cite{Rizzo2010} suggest that these values could be reached with new laser facilities but the very small value of the induced magnetization remains an experimental challenge (Optical field induced magnetization as small as $10^{-10}$ T are reported in \cite{Kalugin2001}).

Symmetry considerations already applied to direct effects indicate that inverse Kerr effect and inverse magneto electric birefringence effects exist in vacuum as well.

\subsubsection{Optical field induced inverse magneto-electric effects\\}

As in the case of magneto-electric birefringences, inverse effects might also be induced by fields associated to electromagnetic waves. As far as we know, this subject has never been treated in literature.

\subsubsection{Parametric amplification induced by an electrostatic field : photon splitting\\}

\begin{figure}[h]
\begin{center}
  \includegraphics[width=8cm]{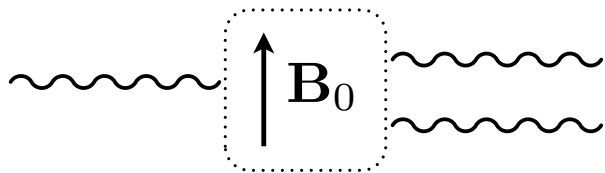}
  \end{center}
  \caption{Parametric amplification induced by an electromagnetic field : splitting of  photon into two photons in the presence of a magnetic field. }\label{photsplit}
\end{figure}

Since the seventies (\cite{Bialynicka-Birula1970}, \cite{Adler1970}) physicists have studied a phenomenon called photon splitting which is the splitting of a photon propagating in vacuum into two photons in the presence of a transverse magnetic field (fig. \ref{photsplit}).
In terms of nonlinear optics textbooks this phenomenon is a parametric amplification induced by an electromagnetic field, but as far as we know it has not been observed in standard media. The incoming photon beam can be considered as the pump electromagnetic wave, and the outcoming photon beams as the signal electromagnetic wave and the idler electromagnetic wave \cite{Shen}. Reference paper for vacuum effect is ref. \cite{Adler1971}. The results of this work have been confirmed more recently in ref. \cite{Adler1996}.

Calculations reported in ref. \cite{Adler1971} are far from being straightforward and they cannot be resumed here easily. We resume in the following the main results and we give some numerical estimations of such an effect.

\begin{figure}[h]
\begin{center}
  \includegraphics[width=8cm]{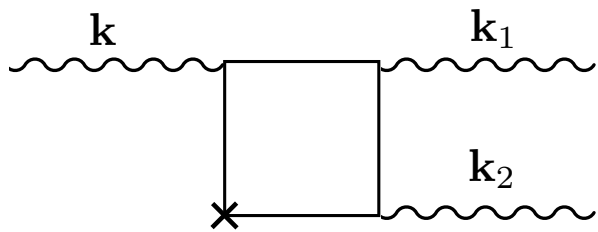}
  \end{center}
  \caption{Box diagram for photon splitting with pump, signal and idler waves all collinear. The $\times$ denotes an interaction with the external field. The contribution of this diagram vanishes in vacuum. }\label{splitbox}
\end{figure}

First of all, let's consider the case of a constant and spatially uniform external magnetic field, with pump, signal and idler waves all collinear. The contribution of the box Feynman diagram corresponding to the interaction of three photons with external field once  (figure \ref{splitbox}) vanishes, so the leading diagram which contributes to photon splitting is the hexagonal one corresponding to three photons interacting with the external field three times (figure \ref{hexsplit}). The pump field attenuation during propagation will be proportional to $B_0^3$, while energy attenuation will be proportional to $B_0^6$. The splitting phenomenon also depends on photon polarization with respect to the external field component perpendicular to the wavevector. Kinematic selection rules and CP selection rules allow only photons polarized parallel to the transverse external field to split, and they can only split into two photons polarized perpendicular to the transverse external field. In principle, photon splitting would polarize an incoming beam in the direction perpendicular to the transverse external magnetic field. This can be summarized in the following formula:

\begin{figure}[h]
\begin{center}
  \includegraphics[width=8cm]{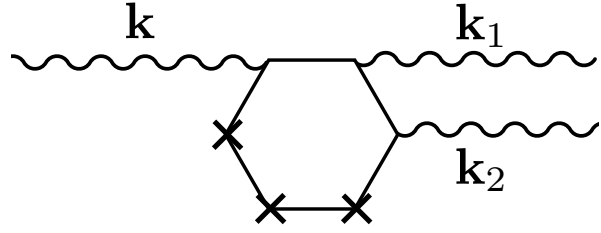}
  \end{center}
  \caption{Hexagon diagram for photon splitting. Each $\times$ denotes an interaction with the external field. }\label{hexsplit}
\end{figure}

\begin{equation}\label{PHsplit}
    \kappa_{(\parallel)\rightarrow(\perp)+(\perp)}\approx\left(\frac{13 \alpha^3}{9450\pi^2}\right)\left(\frac{\hbar\omega}{m_e c^2}\right)^5\left(\frac{B_0 \sin \theta}{B_{cr}}\right)^6\left(\frac{m_e c}{\hbar}\right),
\end{equation}

where $\kappa$ is the attenuation coefficient in m$^{-1}$ and $\theta$ the angle between the external magnetic field and the propagation wavevector.

This formula can be written as
\begin{equation}\label{PHsplitEH}
    \kappa_{(\parallel)\rightarrow(\perp)+(\perp)}\approx\left(\frac{273 \alpha }{8}\right)\left(\frac{\hbar\omega}{m_e c^2}\right)^5\left(\frac{B_{cr}^2}{\mu_0}\right)\left(\frac{B_0 \sin \theta}{\sqrt{\mu_0}}\right)^6\left(\frac{m_e c}{\hbar}\right)c_{3,0}^2
\end{equation}

and it can be further reduced to

\begin{equation}\label{PHsplitred}
    \kappa_{(\parallel)\rightarrow(\perp)+(\perp)} \approx 12.0\left(\frac{\hbar\omega}{m_e c^2}\right)^5\left(\frac{B_0 \sin \theta}{B_{cr}}\right)^6.
\end{equation}

For photons of $\hbar\omega =$ 1.6 $\times$ 10$^{-19}$ J (1 eV) under a 10 T transverse field, $\kappa$ is as small as 4.8 $\times$ 10$^{-80}$ m$^{-1}$.

Taking into account higher order diagrams does not change much this result since their contributions are even smaller than the one from the hexagonal diagram as it has been proved in ref. \cite{Adler1971}. Taking into account radiative corrections should give a correction of the order of $\alpha$, a calculation which has not yet been performed.

\begin{figure}[h]
\begin{center}
  \includegraphics[width=8cm]{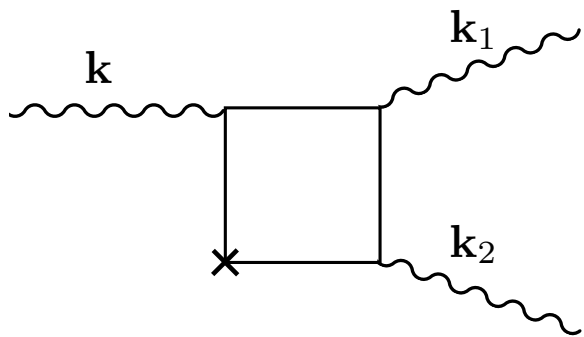}
  \end{center}
  \caption{Box diagram for photon splitting with pump, signal and idler waves non collinear. The $\times$ denotes an interaction with the external field. The contribution of this diagram no more vanishes in vacuum. }\label{photsplitNC}
\end{figure}

If the outcoming photons are no more collinear to the incoming photon the box diagram no longer vanishes (figure \ref{photsplitNC}). This is the case when the magnetic field is no more spatially uniform. This means that the magnetic field can transfer a momentum $p$ that one can write as $\frac{h}{B_{cr}}\frac{\partial B}{\partial l}$ where $l$ is along the direction parallel to the incoming wavevector.
Following ref. \cite{Adler1971}, the ratio between the attenuation coefficient $\kappa_{box}$ due to the box diagram over the attenuation coefficient $\kappa_{hexagon}$ due to the hexagonal diagram can be written as

\begin{equation}\label{ratioK}
   \frac{ \kappa_{box}}{\kappa_{hexagon}}\sim\left(\frac{B_{cr}}{B_0 sin\theta}\right)^4\left(-2\frac{cp}{\hbar\omega}\right)^2.
\end{equation}

If one has a 10 T transverse component and a 1000 T/m magnetic field gradient, $\kappa_{box}$ contribution can be as $10^{10}$ times greater than the $\kappa_{hexagon}$ contribution but the resulting attenuation coefficient would anyway be of the order $10^{-70}$ m$^{-1}$, which is unmeasurable in a terrestrial laboratory. When $B_0$ approaches $B_{cr}$, $\kappa_{hexagon}$ becomes more important than $\kappa_{box}$.

A way to increase photon splitting probability is to use very high fields like the one existing in atoms. The probability increases even more if the photon energy also exceeds $m_e c^2$. As reported in ref. \cite{Lee2003}, a review on photon splitting in atomic fields, Heisenberg Euler lagrangian can only be used to derive photon splitting cross section when $\hbar \omega \ll m_e c^2$. In this limit, total cross section in the coulomb field is determined by the box diagram contribution since the field gradient is important, and can be written in barns ($10^{-28}$ m$^2$) as

\begin{equation}\label{PhSplAt}
   \sigma_{PSAF} = 7.4 \times 10^{-28} \left(\frac{Z^2 \alpha^5}{m_e^2 c^4}\right)\left(\frac{\hbar \omega}{m_e c^2}\right)^6 = 3.7 \times 10^{-25}Z^2\alpha\left(\frac{\omega}{c}\right)^6c_{2,0}^2
\end{equation}

or

\begin{equation}\label{PhSplAtRed}
   \sigma_{PSAF}=2.3 \times 10^{-12}Z^2\left(\frac{\hbar \omega}{m_e c^2}\right)^6,
\end{equation}

where $Z$ is the atomic number.
Such a cross section for an optical photon of energy about 1.6 $\times$ 10$^{-19}$ J, assuming $Z=100$, is about $10^{-24}$ m$^2$ which seems not measurable even using very intense laser beams.

Using high energy photons changes the experimental perspective. The already cited review \cite{Lee2003} gives a clear overview of the field. In particular one can find details about the reported first observation of photon splitting in an atomic field for photons of energy between 120 and 450 MeV \cite{Akhmadaliev2002}. With photons of this energy one gains about 18 orders of magnitude in the cross section with respect to the one for optical photons. Authors of ref. \cite{Akhmadaliev2002} observed about 400 photon splitting events for 1.6 $\times$ 10$^{9}$ incident photons on a Bi$_4$Ge$_3$O$_{12}$ (BGO) target. This result is at 1.6 standard deviations from the prediction of a Monte Carlo simulation based on the most precise QED cross section calculated taking into account the contributions of terms of all orders in the parameter $Z\alpha$. If one uses cross sections calculated only at the lowest orders the difference between simulation and experiment becomes 3.5 standard deviations. The experiment was conducted at the VEPP-4M collider at the Budker Institue of Nuclear Physics, Novosibirsk, Russia \cite{BudInst}. As far as we know, this is still the only observation of such a phenomenon.

\subsubsection{Optical field-induced photon splitting\\}

\begin{figure}[h]
\begin{center}
  \includegraphics[width=8cm]{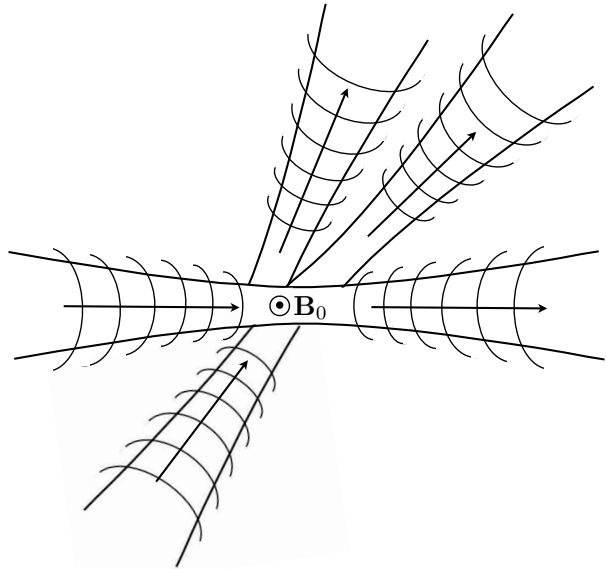}
  \end{center}
  \caption{Optical induced photon splitting : photon splitting is induced by the electromagnetic field generated by an intense electromagnetic wave. }\label{optindPS}
\end{figure}

As for the case of birefringences, one may study the case when photon splitting is induced by an the electromagnetic field generated by an intense electromagnetic wave (see fig. \ref{optindPS}). Theoretical studies of such a phenomenon have been reported in ref. \cite{Affleck1987} and more recently in ref. \cite{DiPiazza2007PRA}.

Photon splitting probability depends essentially of two Lorentz-invariant parameters \cite{DiPiazza2007PRA}, $\eta=\frac{\hbar^2\omega \omega_L}{m_e^2c^4}$ and $\chi=\frac{\hbar\omega E_\omega}{m_ec^2 E_{cr}}$,  where $\hbar\omega$ is the energy of the incoming photon, $\hbar\omega_L$ the energy of a photon of the electromagnetic wave generating the field and $E_\omega$ the electric field associated to this electromagnetic wave.
The case corresponding to $\eta \ll 1$ and $\chi \ll 1$ can be solved using the Heisenberg-Euler lagrangian. The conversion rate obtained is $W_{PSLA}\propto \alpha^3\chi^6$ as shown in ref. \cite{Affleck1987} and \cite{DiPiazza2007PRA}.

Rates in more general cases are given in ref. \cite{DiPiazza2007PRA}. Following conclusions of the authors, it looks that even with fields $E_\omega$ approaching $E_{cr}$ as the ones provided by very powerful future lasers, the conversion rate is such that an observation of optical field induced photon splitting remains very difficult. For example, let's assume that the field is provided by a laser of intensity 10$^{29}$ W.m$^{-2}$ corresponding to $E_\omega \sim$ 5 $\times$ 10$^{-3} E_{cr}$, and that photon energy is 1.6 $\times$ 10$^{-19}$ J,  pulse duration 10 fs, pulse repetition rate 1 Hz. Let's imagine also to have incoming photons of energy between 120 and 450 MeV like in ref.\cite{Akhmadaliev2002} for a total flux of 10$^8$ photons per second.  Following ref. \cite{DiPiazza2007PRA} this experimental configuration would give an event rate smaller than 6 $\times$ 10$^{-4}$ s$^{-1}$. A rate that looks 10$^6$ times smaller than the one corresponding to the observation of photon splitting in the atomic field \cite{Akhmadaliev2002}.

\subsubsection{Second harmonic generation induced by an electromagnetic field : photon fusion\\}

Second harmonic generation and parametric amplification are one the inverse effect of the other. If one exists in vacuum the other exists as well.
The same selection rules apply to both of them. This means that the contribution of the box diagram to the probability of conversion of two collinear incoming photons into one outcoming photon of twice the energy of one incoming photon in the presence of a transverse magnetic field is zero \cite{Raizen1990}, \cite{Ford1990}. At the lowest orders the transition probability is therefore proportional to $\left(\frac{B_0}{B_{cr}}\right)^6$ and as for the photon splitting, photon fusion of collinear photons in a uniform magnetic field looks unmeasurable in any terrestrial laboratory. As far as we know, second harmonic generation induced by a magnetic field has never been observed even in standard media, while second harmonic generation induced by an electric field has been observed since the seventies \cite{Bethea1975}.

The box diagram does not vanish if the incoming photons are not collinear and/or the magnetic field is not uniform. These cases are treated in ref. \cite{Ding1992} and more recently in ref. \cite{Kaplan2000}. In particular, the authors of this two papers treated the example of a pulsed plane wave or a Gaussian laser beam propagating in uniform or non uniform dc fields. Following ref. \cite{Kaplan2000}, an order of magnitude of the expected photon rate for a gaussian beam in a uniform magnetic field can be obtained by the following formula

\begin{equation}\label{PhSplLa}
   N_{SHG}=1.2 \times 10^{-36} \rho I \lambda B_0^2,
\end{equation}

where the numerical factor is proportional to $c_{2,0}^2$, $\rho$ is the total time-averaged power of fundamental harmonics, $\lambda$ is the wavelength of fundamental harmonic, $I$ is the laser maximal intensity at focal point. Taking as values $\rho=10^5$ W, $I = 10^{26}$ W.m$^{-2}$, $\lambda=0.8$ $\times$ 10$^{-6}$ m, as the authors of ref. \cite{Kaplan2000}, and $B_0=50$ T one obtains $N_{SHG}\sim 2.5$ $\times$ 10$^{-8}$ s$^{-1}$ which seems very difficult to observe.
For sake of argument let's recall that best single photon detectors in the near infrared region has dark count rates of the order of 10$^{-3}$ s$^{-1}$ (see e.g. \cite{Miller2003}).

\subsubsection{Optical field induced photon fusion\\}

Photon fusion should also be induced using an electromagnetic wave as the field source but no paper about it exists in literature, as far as we know. As for the case of photon splitting one may guess that the optical field induced effect should not be easier to be observed than the one induced via electrostatic fields.

\subsubsection{Intensity dependent refractive index\\}

As already discussed, the index of refraction for light propagating in the presence of electromagnetic field depends on the field amplitude. This means that if the field is not uniform but it depends on the spatial coordinates light will propagate in the presence of a gradient of the index of refraction. It is also known that light ray generally bends towards regions with a higher index of refraction. In this case the ray behavior can be described solving what is called the eikonal equation \cite{BornWolf}. For example, light passing in a vacuum near a magnetic pole will bend toward the pole giving the impression to be attracted by the magnetic pole itself. A rough estimation of the deviation angle $\theta_d$ is $n(B_0)/n(0)$ which for vacuum reduces to $\theta_d \sim 7 c_{2,0} \frac{B_0^2}{\mu_0}\approx 9.3 \times 10^{-24}B_0^2$ for light polarized parallel to $\bf{B}_0$.

The magnetic deviation calculated using eikonal equation in the case of magnetic dipole has been given in ref. \cite{Denisov2001}, and \cite{Dupays2005}. It depends on the magnetic moment orientation with respect to the light propagation direction, and it scales with the minimal distance of the light ray to the magnetic moment $\rho_m$ as $1/\rho_m^6$. Our rough estimation holds also for this more complicated case.

Actually an experiment has been performed around 1961 to look for such a deviation and it has been reported in ref. \cite{Jones1961}. The maximum field was of about 1 T and results indicated that the magnetic deviation was less than 5 $\times$ 10$^{-13}$ rad which is obviously in agreement with our estimation of about 9 $\times$ 10$^{-24}$ rad.

It is clear that if one imagines a powerful laser beam, such that during propagation energy density corresponds to fields approaching the critical ones, self focusing should be observed because external rays will bend towards the inner region of the beam where the index of refraction is higher. A discussion of this effect and other related to it can be found in ref. \cite{Soljacic2000}. From the experimental point of view the problem is that to get fields of the order of the critical ones one needs lasers with a power of the order of 10$^{33}$ W.m$^{-2}$ and this will may be possible only in a far future since ongoing projects have target intensity of 10$^{30}$ W.m$^{-2}$ \cite{ELI}.

The QED vacuum effects at the interfaces between region of different $n$, like the creation of evanescent waves, has also been studied in ref. \cite{Li2011}.

Let's imagine now following ref. \cite{DiPiazza2006} proposal a x-ray probe beam traversing a standing electromagnetic wave. In any point of the standing wave one can define an electromagnetic field and therefore an index of refraction that depends on spatial coordinates. In principle the standing wave acts on the probe beam as a diffraction grating and a diffraction by a standing wave should be observed. The calculation of the nonlinear phase shift acquired by crossing electromagnetic waves in a vacuum is also reported \cite{Ferrando2007}. A more recent proposal to observe light by light diffraction in vacuum can be found in ref. \cite{Tommasini2010PRA}. A natural extention of such a kind of apparatus is to have more than a standing wave to observe double-slit light by light interference as proposed in ref. \cite{King2010}, and ref. \cite{King2010NP} or strong periodic fields structured \cite{Yu2011} to induce Bragg scattering \cite{BraggS}. Authors of ref. \cite{King2010} argue that for experimental parameters attainable at upcoming XFEL and at ELI facilities (80 GW of x-rays in a 100 fs pulse of 0.4 nm wavelength focused in a spot of 100 $\mu$m radius, and ELI expected power focused in a diffraction limited spot), approximately 2 photons per laser shot can be diffracted so that to give a measurable signal.

\subsubsection{Photon-Photon scattering\\}

Since the original paper of Euler and Kochel \cite{Euler1935} it was clear that in the framework of Quantum ElectroDynamics photon-photon scattering was allowed in vacuum. First determination of the photon-photon scattering amplitude can be found in ref. \cite{Euler1935}. This result has been confirmed later using Feynman diagrams \cite{Karplus1951}. Very recently, a tutorial paper showing how to compute low-energy photon-photon scattering has also been published \cite{Liang2012}.

In the low energy approximation $\hbar\omega \ll m_e c^2$ and for unpolarized light, the total cross section can be written as

\begin{equation}\label{PhPhsca}
   \sigma_{\gamma\gamma\rightarrow\gamma\gamma}=\frac{973}{10125\pi}\alpha^2 r_e^2 \left(\frac{\hbar\omega}{m_e c^2}\right)^6=\frac{973}{20\pi}\left(\frac{(\hbar\omega)^6}{\hbar^4 c^4}\right)c_{2,0}^2
\end{equation}

where the electron classical radius is $r_e= \alpha \mathchar'26\mkern-10mu\lambda$, and $\hbar\omega$ is the energy of a photon in the reference frame of the center of mass \cite{PCM}. 

This means that $\sigma_{\gamma\gamma\rightarrow\gamma\gamma} \approx 7.3 \times 10^{-70}$ m$^2$ when $\hbar\omega = 1.6$ $\times$ 10$^{-19}$ J. 

The cross section increases very rapidly with photon energy, reaches a maximum of $1.6 \times 10^{-34}$ m$^2$ when $\hbar\omega = 1.5$ $m_e c^2$ and then it decreases as $(1/\hbar\omega)^2$ \cite{DeTollis1964} and \cite{Dicus1998}.

As far as we know the first attempt to observe photon-photon scattering dates from 1928 \cite{Vavilov1928} (see also ref. \cite{Vavilov1930}), followed in 1930 by the experiment reported in ref. \cite{Hughes1930}. No scattered light was detected corresponding to an upper limit $\sigma_{\gamma\gamma\rightarrow\gamma\gamma} \leq 3 \times 10^{-24}$ m$^2$.
Notwithstanding proposals to use gamma rays \cite{Harutyunian1964} or X rays \cite{Rosen1965} to take advantage of the higher cross section at these energies, no other experiment has been tried until 1996, when a new attempt to observe photon-photon scattering is reported in ref. \cite{Moulin1996}. This experiment was based on two laser beams provided by LULI \cite{LULI}, France, colliding head-on in vacuum with a center of mass energy of 1.7 eV. The corresponding QED cross section was 1.6 $\times$ 10$^{-68}$ m$^2$. No scattered photon was observed and authors gave an upper limit of the cross section $\sigma_{\gamma\gamma\rightarrow\gamma\gamma} \leq 9.9 \times 10^{-44}$ m$^2$.

\begin{figure}[h]
\begin{center}
  \includegraphics[width=6cm]{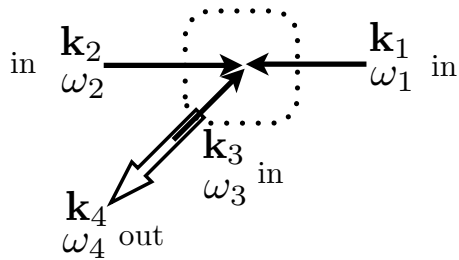}
  \end{center}
  \caption{Photon photon scattering in four wave mixing configuration. The collision of three beams satisfying the resonance condition in vacuum $\bf{k_1}+\bf{k_2}=\bf{k_3}+\bf{k_4}$ and $\omega_1+\omega_2=\omega_3+\omega_4$, generate a fourth beam in the direction given by the matching conditions. The generated beam is stimulated by the third beam. This configuration avoids the detection of scattered incoming photons from the other beams.}\label{photonscat}
\end{figure}

In ref. \cite{Dewar1974} photon scattering is studied in the case of three incoming beams arranged so that the energy and momentum conservation conditions $\hbar(\omega_1+\omega_2-\omega_3)=\hbar\omega_4$ , $\bf{k_1}+\bf{k_2}-\bf{k_3}=\bf{k_4}$ can be satisfied by a fourth, scattered beam. This configuration stimulates the photon emission in the fourth beam and so the matching conditions fix the direction of the generated wave (see fig \ref{photonscat}).

This three beam geometry is also discussed in ref. \cite{Grynberg1990}. Following these proposals, a new experiment has been mounted at LULI \cite{LULI}, France, and results reported in ref. \cite{Bernard2000} in 2000.

The chosen geometry was the following : $\bf{k_1}$=k\textbf{\^ x}, $\bf{k_2}$=k\textbf{\^ y}, $\bf{k_3}$=$\frac k 2$ \textbf{\^ z} and $\bf{k_4}$=k\textbf{\^ x}+ k\textbf{\^ y}-$\frac k 2$ \textbf{\^ z}. Authors have chosen wavelength of 800 nm for beams 1 and 2, and 1300 nm for the third one. The generated signal was expected in the visible at 577 nm.
Thanks to this geometry, both  the wavelength and the direction of the generated wave is different from the others, allowing to easily separate photons from QED scattering from photons belonging to the incoming lasers. Energy of the two main beams at 800 nm was 0.4 J with a duration of 40 fs (Full Width of Half Maximum), while the third one is generated in an optical amplifier to shift its wavelength up to 1300 nm. The three beams are injected in a vacuum chamber with a diameter of 3 cm. Authors have calibrated their apparatus by measuring the third order susceptibility of nitrogen gas.  In vacuum no evidence for photon-photon scattering was observed and authors gave an upper limit of the cross section $\sigma_{\gamma\gamma\rightarrow\gamma\gamma} \leq 1.5 \times 10^{-52}$ m$^2$, at 18 orders of magnitude from QED prediction.

In principle vacuum can be treated as a standard medium and, as shown in ref. \cite{Moulin1999} a third-order nonlinear effective susceptibility $\chi_v^{(3)}$ can be defined

\begin{equation}\label{ChiV}
   \chi_v^{(3)}=\frac{K}{45\pi\alpha}\left(\frac{r_e e}{m_ec^2}\right)^2= 2\epsilon_0 K c_{2,0}.
\end{equation}

$K$ is a factor that depends on the directions of the incident beams and of their polarization $(K<14)$ \cite{Moulin1999}. For $K=1$, one gets $\chi_v^{(3)}\approx 3.0 \times 10^{-41}$ m$^2$.V$^{-2}$.

In recent years several new proposals have been published but no new experiment has been performed. While in ref. \cite{Brodin2001} the possibility to measure photon-photon scattering using electromagnetic modes in a waveguide is discussed, the other proposals \cite{Lundin2006}, \cite{Lundstrom2006}, \cite{Fedotov2007}, \cite{Tommasini2008}, \cite{Tommasini2009}, \cite{Tommasini2010}  are mostly motivated by the recent evolutions in the field of very powerful laser sources which constitute a very important tool for fundamental physics \cite{Marklund2009}, \cite{Heinzl2009}.  In particular the perspectives of photon-photon interaction experiments using high intensity laser are discussed in ref. \cite{Marklund2006RMP}.

The connection between the vacuum index of refraction $n$ in the presence of an external electromagnetic field and photon-photon scattering amplitude in the forward direction $f_0$ can be established thanks to the optical theorem \cite{Haissinski2006}

\begin{equation}\label{OpTh}
   n=1+\frac{2\pi}{k^2}Nf_0,
\end{equation}

where $N$ is the average density of centers of scattering that is proportional to the energy density of the external field and inversely proportional to the photon energy in the center of mass reference frame.

Photon-photon cross section in the low energy limits depends on the lowest order coefficients of the development of the Heisenberg-Euler lagrangian like $c_{2,0}$. This is true for any vacuum effect calculated in the same approximation and therefore any experiment testing one of these effect and therefore measuring such coefficients also indirectly test the existence of all the other effects. For example authors of ref. \cite{Bregant2008} argue that attempts to measure vacuum magnetic birefringence give better indirect limits on $\sigma_{\gamma\gamma\rightarrow\gamma\gamma}$ than the one obtained by experiment reported in ref. \cite{Bernard2000}.

\subsection{Vacuum dichroism}

In 1964 \cite{Klein1964} was pointed out that the Heisenberg-Euler lagrangian has imaginary contributions due to the poles in eq. \ref{LHE}. This means that together to a real part of the index of refraction which gives rise to vacuum electromagnetic birefringences it exists an imaginary part which corresponds to vacuum dichroism i.e. absorption of photons in vacuum depending on photon polarization. This absorptive index of refraction arises for the case of a pure external electric field. For the case of a pure magnetic field there is no absorptive contribution to the index of refraction. If a static external electric field $E_0$ is present, one can write \cite{Klein1964}

\begin{equation}
n_{\bot} \simeq  1
\end{equation}

and

\begin{equation}
n_{\|} \simeq 1 + \frac{1}{4}i \alpha\frac{1}{e^{\frac{\pi E_{cr}}{E_0}}-1},
\end{equation}

where it is clear that only light polarized parallel to the external field can be absorbed. Finally a linear absorption coefficient $\kappa_{\|}$ can be defined

\begin{equation}
\kappa_{\|} =  \pi\frac{\alpha}{\lambda}\frac{1}{e^{\frac{\pi E_{cr}}{E_0}}-1}
\end{equation}

In principle, the behavior of the vacuum in a strong electric field is such that an unpolarized light beam becomes partially polarized because the component of the electric vibration parallel to the electric field is absorbed whereas the other component is not. The electric field required to observe such a vacuum dichroism is of the order of $E_{cr}$.

With the advent of very powerful laser sources, fields approaching the critical ones can be achieved, and vacuum dichroism will eventually be observed by the associated pair production by laser field \cite{Dunne2009} (see also \cite{Labun2011} and refs. within).

\subsection{Delbr\"{u}ck scattering}

Electric fields as high as the critical one can be found at the surface of atomic nuclei. To probe such a field light has to have wavelength of the order of the nucleus radius which means high energy photons.

When high energy photons pass by a nucleus, they can not only be absorbed but they can also be deflected by the Coulomb field. This phenomenon is known as the Delbr\"{u}ck scattering since Delbr\"{u}ck first proposed it \cite{Delbruck1933} to explain the results of an experiment in which 2.615 MeV photons were scattered by lead and iron \cite{Meitner1933}. From a phenomenological point of view, in the low energy limit this effect can be viewed as a consequence of the refractive index induced by the Coulomb field. As shown in ref. \cite{Rohrlich1952}, in the high energy limit Delbr\"{u}ck scattering can be related to pair production via the optical theorem. As discussed in reviews \cite{Milstein1994} and \cite{Schumacher1999} comparison between scattering experimental data and theoretical prediction of Delbr\"{u}ck scattering are complex because the measured total scattering cross section depends also on the contribution of the Rayleigh scattering and nuclear Compton scattering. The first clear observation of Delbr\"{u}ck scattering was reported by Schumacher et al. \cite{Schumacher1975} using 2.754 MeV photons on lead nuclei. The energy range around 2.7 MeV is somewhat ideal to observe Delbr\"{u}ck scattering since at lower energy Rayleigh scattering dominates, while at higher energy Compton scattering dominates. Experimental data cannot be explained without taking into account the Delbr\"{u}ck scattering while to exactly reproduce their value further corrections to the theoretical predictions, called Coulomb corrections, have to be considered. The most recent measurements of high-energy Delbr\"{u}ck scattering \cite{Akhmadalev1998} have been performed at the Budger Institute of Nuclear Physics  \cite{BudInst}, Novosibirsk, Russia.

\section{Phenomenology in astrophysics}

As we have discussed in all the previous paragraphs, vacuum non linearities need to be detected an intense electromagnetic radiation source and strong electromagnetic fields. This demand is a real challenge in terrestrial laboratories but there are places in the cosmos where both radiation and fields exist. The celestial bodies that have attracted most the attention of scientists to observe vacuum effects are strongly magnetized stars, neutron stars \cite{Harding2006} and white dwarfs \cite{Koester2002}. Both kind of celestial objects are the remnants of the implosion of massive stars: stars of a mass exceeding the solar one in the case of neutron stars, and stars of masses about the solar one for white dwarfs. Neutron stars have magnetic fields on their surface typically of about 10$^7$-10$^8$ T but special neutron stars, called magnetars, are supposed to have surface magnetic fields exceeding $B_{cr}$. White dwarf field typically ranges from 50 T to 10$^4$ T. The explanation of such high fields is that during implosion the magnetic field flux is conserved. A star like the sun has a field of about 5 $\times$ 10$^{-3}$ T and a radius of about 7 $\times$ 10$^5$ km, and after implosion a neutron star as a radius of about 10 km and a white dwarf of about 10$^4$ km. The surface magnetic field is therefore much higher. Stars are obviously sources of a large spectrum of radiation and the implications of QED in the observed emission of such celestial objects is a very wide and important field that has been already treated in specific reviews like in ref. \cite{Harding2006} and more recently in ref. \cite{HeylCargese}.

Magnetic ray bending in neutron stars has been studied in ref. \cite{Denisov2001},  \cite{Dupays2005} and  \cite{Cuesta2006}. This phenomenon is related to the fact that the refractive index depends on the field intensity. Like in the 1961 terrestrial experiment already cited \cite{Jones1961}, the idea consists in observing the deflection of a light ray passing in a region where a non uniform magnetic field is present. In ref. \cite{Dupays2005} a specific neutron star binary system is considered. Both stars are pulsars which means that they emit a radiation beam along their magnetic dipole direction. Periodically the beam of one of the two stars is eclipsed by the other one thus periodically the radiation beam of one of the two stars traverses a region where a field of several Tesla exists over a distance of several hundreds of kilometers and it should be bent in such a way to give to the signal observed on the earth a clear signature of this QED magnetically induced lensing. This kind of lensing effect is similar to the more studied gravitational one where light is bent by the gravitational field \cite{Bartelmann2001}.

Magnetically induced spontaneous vacuum energy emission via the production of real electron-positron pairs, which is also called vacuum breakdown, and its implication for neutron star emission is the subject of ref. \cite{Xue2003}. The basic idea is that it is possible to release vacuum-energy from the vacuum by introducing an external magnetic field because of the difference of vacuum energy with and without the field. In the case of black holes \cite{BHs}, which are the remnants of the explosion of very massive stars, vacuum breakdown in an external field has also been studied \cite{Treves1999}.

QED vacuum magnetization around neutron stars and its effect on neutron star braking is studied in ref. \cite{Dupays2008} and \cite{Dupays2012}. This phenomenon, that the authors call quantum vacuum friction, consists in the interaction between the star magnetic dipole and the magnetic dipole induced in vacuum because of vacuum magnetization predicted in the framework of QED. The star rotates around its own axis thus the vacuum dipole also change in time. The two are not collinear because vacuum dipole, which is the sum of the magnetization in all the points around the neutron star, is due to a retarded field since the velocity of light is not infinite. The interaction between the two dipoles give rise to a couple that tends to brake the star. This has important consequences on the spindown evolution of neutron stars in particular in the case of magnetars.

The possibility to detect QED effects on the polarization of radiation emitted by neutron star and white dwarfs is discussed in \cite{Heyl2002}, \cite{Lai2003}, \cite{Potekhin2004},  \cite{Kerkwijk2004},  \cite{Shannon2006}, \cite{Gnedin2006}, \cite{vanAdelsberg2008}, \cite{Wang2009}.
Combined effects of QED vacuum and plasma in the neighborhood of a neutron star are studied in \cite{Heyl1999}, \cite{DiPiazza2007}, \cite{Brodin2007PRL}. Photon-plasma interactions are also the object of a review \cite{Marklund2006RMP}. In general radiation emitted from the star propagates in a plasma but also in the presence of high magnetic fields. Light polarization changes during propagation so that star atmosphere emits polarized light even if the star surface radiation is not polarized. The emission appears also to vary across the surface. The polarization across the surface should also show special patterns because the rotating magnetic field twists the polarization. All these subtle effects gives a clear signature to radiation observed on the earth.

Nonlinear propagation of electromagnetic waves in the magnetosphere of a magnetar, taking into account both the QED vacuum and the magnetized plasma, has been also studied non-perturbatively in ref. \cite{Mazur2011}. The nonlinear behaviour of these electromagnetic waves should play an important role in the energy transmission in pulsars and magnetars.

Since the very beginning magnetized stars have been considered as the best choice to observe photon splitting (\cite{Bialynicka-Birula1970}, \cite{Adler1970}). In a star magnetosphere only splitting of a photon polarized parallel to the field into two photons polarized perpendicular to the field is allowed because the index of refraction of the parallel propagation mode is greater than the perpendicular propagation mode. This is the only process that conserves momentum. Again, magnetized stars like neutron stars should appear as emitting polarized light. Moreover, photon splitting is more efficient for high energy photons thus photon splitting has often been invoked to explain hard cutoffs in gamma-ray spectra from pulsars  \cite{Baring2001}, but data are difficult to interpret because one does not know where gamma-rays are actually produced.

All this wide research domain has triggered the construction of a special satellite GEMS \cite{GEMS} which goal is to measure the polarization of X rays sources. Its launch is scheduled after November 2014, and neutron stars are between its prime targets.

\section{Non trivial vacua}

Examples of what is called non trivial vacua can be found in literature. This essentially means that light propagates in a vacuum in which a distribution of real or virtual photons exists. In this case one can calculate an average value for the $E^2$ and $B^2$ and one can use this value to calculate the corresponding effects due in particular to four wave mixing in vacuum. For example the velocity of light propagating perpendicular to two parallel infinite conducting plates and in the region between them has been first calculated by Scharnhorst \cite{Scharnhorst1990}. The variation of the velocity of light is due to virtual photons energy density which is also responsible of the well known Casimir effect \cite{Casimir1948}. The conceptual problem is that in the case treated by Scharnhorst the velocity of light should exceed $c$. This has attracted a lot of interest (see e.g. ref. \cite{Milonni1990}), but as far as we understand, this effect is also expected to be undetectable with up-to-date apparata. The effect of real photon radiation associated to temperature with comparison to the one associated to virtual photons is also discussed in ref. \cite{Dittrich1998}.

In the case of very intense photon beams, the photon density is so high that in literature the term of radiation gas is used to indicate such an environment. QED photon-photon interactions change the refractive index of a radiation gas and new optical phenomena should appear. Selffocusing, the formation of so-called photon bullets \cite{Brodin2003}, \cite{Marklund2004}, wave collapse \cite{Shukla2004PRL}, \cite{Marklund2005} and a phenomenon called photon acceleration in vacuum \cite{Mendonca2006}.

\section{CP violating vacuum}

In the framework of the standard model, CP violation has been observed (see e.g. \cite{Aubert2008}). In principle CP violation in photon-photon interactions would modify the quantum vacuum behavior also in the case of light propagation. In particular terms corresponding to coefficients $c_{0,1}$ and $c_{2,1}$ could exist in the lagrangian $L$. The phenomenology of the existent of a term proportional to $F^2G$ has been studied in ref. \cite{Pinto2006} and \cite{Millo2009}.

One of the main results is that a new type of Jones Birefringence depending on $B_0^2$ should exist. Nevertheless the predicted effect in the framework of the standard model is very small compared to the QED birefringence effects because the energy scale of such a new lagrangian term is the QCD one \cite{Millo2009}.

\section{Conclusion}

From the phenomenological point of view quantum vacuum can be treated as a standard nonlinear optical medium. In the low photon energy limit and for fields smaller than the critical ones the lowest order terms of the Heisenberg-Euler lagrangian are sufficient to give precise predictions of the effects to be measured like in the case of vacuum Cotton-Mouton magnetic birefringence. In principle, higher $\alpha$ contributions could be calculated giving predictions more and more precise to compare with experimental results as in the case of other important QED tests as for example the anomalous magnetic moment of the electron (see e.g. ref. \cite{Gabrielse2006}). Our review is mainly devoted to this low energy, small field limit. It is worth stressing that vacuum Cotton-Mouton magnetic birefringence is the field where the only ongoing experiments exist. Experimentalists are at three orders of magnitude from the QED predictions \cite{Zavattini2012} and there is hope that in few years one of the most fundamental predictions of Heisenberg-Euler lagrangian will be tested experimentally.

At higher photon energies and/or at higher fields, theoretical calculations becomes more difficult since they are no more perturbative in $\alpha$. A large theoretical literature exists covering a very wide spectrum of phenomena. Some of these works have been triggered by the new intense lasers facilities under development.
Actually, intense lasers should give access to new and impressive QED effects (see e.g. the reviews \cite{Heinzl2012} and \cite{DiPiazza2012}) that have been treated in this review but that certainly deserve more space. Last experiments on photon-photon scattering were at 18 orders of magnitude from the QED prediction but a large number of proposals are just waiting for the new facilities to be operational and eventually a complete new field will be opened.

For photons in MeV region, both Delbr\"{u}ck scattering \cite{Schumacher1975} and recently photon splitting in an atomic field \cite{Akhmadaliev2002} have been observed confirming once more that QED is a very powerful tool to describe nature.

Last but not least, astrophysical tests of quantum vacuum properties seem very promising and hopefully, thanks to more and more precise observational data, the cosmos itself will become a laboratory to test QED photon-photon interactions.

\section{Acknowledgements}

We thank Mathilde Fouch\'e and Geert Rikken for strong support and very useful discussions. We also thank Mathilde Fouch\'e for carefully reading the manuscript.


\newpage
\begin{thebibliography} {99}

\bibitem{Aristotle} Aristotle, Physics, book IV, part VI-IX.

\bibitem{Jackson} J.D. Jackson, {\it Classical electrodynamics}, J. Wiley \& Sons.

\bibitem{Faraday1845} M. Faraday, {\it Phil. Mag.} {\bf 28} (1846), 294 ; {\it Phil. Trans. R. Soc.} {\bf 136} (1846), 1.

\bibitem{Franken1961} P.A. Franken, A.E. Hill, C.W. Peters, and G. Weinreich, {\it Phys. Rev. Lett.} {\bf 7} (1961), 118.

\bibitem{Shen} Y.R. Shen, {\it The Principles of Nonlinear Optics}, (Oxford university press 2007) $6^{th}$ ed..

\bibitem{Landau} Landau-Lifschitz, {\it Quantum Electrodynamics}, Mir Edition.

\bibitem{Buckingham} A.D. Buckingham, and J.A. Pople, {\it Proc. Phys. Soc.} {\bf 1369} (1956), 1133

\bibitem{Oppenheimer1933} J.R. Oppenheimer and M.S. Plesset, {\it Phys. Rev.} {\bf 44} (1933), 53.

\bibitem{Dirac1934}
P.A.M. Dirac, {\it Rapport du 7$^e$ Conseil Solvay de Physique, Structure et Propriet\'es des Noyaux Atomiques} (1934), 203.

\bibitem{Halpern1933} O. Halpern, {\it Phys. Rev.} {\bf 44} (1933), 855.

\bibitem{Breit1934} G. Breit and J.A. Wheeler, {\it Phys. Rev.} {\bf 46} (1934), 1087.

\bibitem{Euler1935} H. Euler et B. Kochel, {\it Naturwiss.} {\bf 23} (1935), 246.

\bibitem{Euler1936} H. Euler, {\it Annalen der Physik} {\bf 5} (1936), 398.

\bibitem{Heisenberg1936} W. Heisenberg and H. Euler, {\it Z. Phys.} {\bf 98} (1936), 714.

\bibitem{Weisskopf1936} V. Weisskopf, {\it Mat.-Fis. Med. Dan. Vidensk. Selsk.} {\bf 14} (1936), 6.

\bibitem{Bialynicka-Birula1970} Z. Bialynicka-Birula and I. Bialynicki-Birula, {\it Phys. Rev. D} {\bf 2} (1970), 2341.

\bibitem{Libro} S. Schweber, {\it QED and the men who made it}, Princeton University Press.

\bibitem{Jauch1948} J.M. Jauch and K.M. Watson, {\it Phys. Rev.} {\bf 74} (1948), 950.

\bibitem{Karplus1950} R. Karplus and M. Neuman, {\it Phys. Rev.} {\bf 80} (1950), 380.

\bibitem{Schwinger1951} J. Schwinger, {\it Phys. Rev.} {\bf 82} (1951), 664.

\bibitem{Heyl1997} J.S. Heyl, and L. Hernquist, {\it Phys. Rev. D} {\bf 55} (1997), 2449.

\bibitem{Dunne2012} G.V. Dunne, {\it Int. J. Mod. Phys.} {\bf 27} (2012), 1260004.

\bibitem{Ritus1975} V.I. Ritus, {\it Soviet Physics-JETP} {\bf 42} (1975), 774.

\bibitem{Serber1935} R. Serber, {\it Phys. Rev.} {\bf 48} (1935), 49.

\bibitem{Uehling1935} E.A. Uehling, {\it Phys. Rev.} {\bf 48} (1935), 55.

\bibitem{Heyl1997_2} J.S. Heyl and L. Hernquist {\it J. Phys. A: Math. Gen.} {\bf 30} (1997), 6475.

\bibitem{Huttner1992} B. Huttner, and S.M. Barnett, {\it Phys. Rev. A} {\bf 46} (1992), 4306.

\bibitem{Philbin2010} T.G. Philbin, {\it New Journal of Physics} {\bf 12} (2010), 123008.

\bibitem{Shukla2004} P.K. Shukla, M. Marklund, D.D. Tskhakaya, and B. Eliasson, {\it Phys. Plasmas} {\bf 11} (2004), 3767.

\bibitem{Kabat2002} D. Kabat, K. Lee, and E. Weinberg, {\it Phys. Rev. D} {\bf 66} (2002), 014004.

\bibitem{Maiman} T. H. Maiman, {\it Nature} {\bf 187} (1960), 493.

\bibitem{ELI} http://www.extreme-light-infrastructure.eu/

\bibitem{ExaJapan} http://www.ile.osaka-u.ac.jp/

\bibitem{Mourou}  D. Strickland, and G. Mourou, {\it Opt. Commun.} {\bf 56} (1985), 219.

\bibitem{Yanovsky2008} V. Yanovsky, V. Chvykov, G. Kalinchencko, P. Rousseau, T. Planchon, T. Matsuoka, A. Maksimchuk, J. Nees, G. Cheriaux, G. Mourou, and K. Krushelnick, {\it Opt. Express} {\bf 16} (2008), 2109.

\bibitem{HERCULES} http://www.engin.umich.edu/research/cuos/ResearchGroups/HFS/Experimentalfacilities/\\HERCULESPetawattLaser.html

\bibitem{NIF} https://lasers.llnl.gov/

\bibitem{MegaJ} http://www-lmj.cea.fr/

\bibitem{LHCmag} http://public.web.cern.ch/public/en/lhc/howlhc-en.html

\bibitem{NHMFL} http://www.magnet.fsu.edu/

\bibitem{LNCMIG} http://ghmfl.grenoble.cnrs.fr/

\bibitem{HFML} http://www.ru.nl/hfml/

\bibitem{HMFLHef} http://english.hfcas.ac.cn/HMFL.htm

\bibitem{TML} http://www.nims.go.jp/TML/english/

\bibitem{NHMFL-PF} http://www.magnet.fsu.edu/about/losalamos.html

\bibitem{LNCMIT} http://www.toulouse.lncmi.cnrs.fr/

\bibitem{HDL} http://www.hzdr.de/db/Cms?pNid=580

\bibitem{WHMFC} http://eng.whmfc.cn/category/10/2012-01-13/125449343.html

\bibitem{MegaGauss} http://www.toulouse.lncmi.cnrs.fr/spip.php?page=rubrique$\&$id$\_$rubrique=36$\&$lang=en

\bibitem{TrGen} P. Frings, J. Vanacken, C. Detlefs, F. Duc, J.E. Lorenzo, M. Nardone, J. Billette, A. Zitouni, W. Bras, and G.L.J.A. Rikken, {\it Rev. Sci. Instr.} {\bf  77} (2006), 063903.

\bibitem{Crosser2010} M.S. Crosser, S. Scott, A. Clark, and P.M. Wilt, {\it Rev. Sci. Instr.} {\bf 81} (2010), 084701.

\bibitem{LULI} http://www.luli.polytechnique.fr/

\bibitem{Bass1962} M. Bass, P.A. Franken, J.F. Ward, and G. Weinreich, {\it Phys. Rev. Lett.} {\bf 9} (1962), 446.

\bibitem{Yariv} A. Yariv, and P. Yeh, {\it Photonics}, (John Wiley \& Sons, New York 1984) 1st ed.

\bibitem{Giordmaine} J.A. Giordmaine and R.C. Miller, {\it Phys. Rev. Lett.} {\bf 14} (1965), 973.

\bibitem{Pockels1893} F. Pockels, {\it Abhandl. Gesell. Wiss. Gottingen} {\bf 39} (1893), 1 ; F. Pockels, {\it Lehrbuch der Kristalloptik} (Teubner, Leipzig, 1906)

\bibitem{Drung2007} D. Drung, C. Assmann, J. Beyer, A. Kirste, M. Peters, F. Ruede, and Th. Schurig, {\it IEEE Transactions on Applied Superconductivity} {\bf 17} (2007), 699.

\bibitem{Cohen2009} T.D. Cohen, and E.S. Werbos, {\it Phys. Rev. C} {\bf 80} (2009), 015203.

\bibitem{Karbstein2012} F. Karbstein, L. Roessler, B. D\"{o}brich, and H. Gies, {\it Int. J. Mod. Phys. Conf. Ser.} {\bf 14} (2012), 403.

\bibitem{Kerr1875}  J. Kerr, {\it Br. Assoc. Rep.} (1901), 568.

\bibitem{Cotton}
A. Cotton and H. Mouton, {\it Cr. r. hebd. S\'eanc Acad. Sci., Paris} {\bf
141} (1905), 317 and
349 ; {\it Ibid.} {\bf 142} (1906), 203 ; {\it Ibid.} {\bf 145} (1907), 229
;  {\it Ann. Chem. Phys.} {\bf 11} (1907), 145 and 289.

\bibitem{Roth2000} T. Roth and G.L.J.A. Rikken, {\it Phys. Rev. Lett.} {\bf 85} (2000), 4478.

\bibitem{Roth2002} T. Roth and G.L.J.A. Rikken,  {\it Phys. Rev. Lett.} {\bf 88} (2002), 063001.

\bibitem{Ross1989} H.J. Ross, B.S. Sherbone, and G.E. Stedman, {\it J. Phys. B: At. Mol. Opt. Phys.} {\bf 22} (1989), 459.

\bibitem{Rikken2002} G.L.J.A. Rikken, C. Strohm and P. Wyder, {\it Phys. Rev. Lett.} {\bf 89} (2002), 133005.

\bibitem{Rizzo2005} C. Rizzo and G.L.J.A. Rikken, {\it Physica Scripta} {\bf 71} (2005), C5.

\bibitem{Adler1971} S. Adler, {\it Ann. Phys. N.Y.} {\bf 67} (1971), 599.

\bibitem{Denisov2000} V.I. Denisov, {\it Phys. Rev. D} {\bf 61} (2000), 036004.

\bibitem{Rikken2000} G.L.J.A. Rikken and C. Rizzo, {\it Phys. Rev. A} {\bf 63} (2000), 012107.

\bibitem{Bakalov1998} D. Bakalov, F. Brandi, G. Cantatore, G. Carugno, S. Carusotto, F. Della Valle, A.M. De Riva, U. Gastaldi, E. Iacopini, P. Micossi, E. Milotti, R. Onofrio, R. Pengo, F. Perrone, G. Petrucci, E. Polacco, C. Rizzo, G. Ruoso, E. Zavattini, and G. Zavattini, {\it Quantum Semiclass. Opt.} {\bf 10} (1998), 239.

\bibitem{CODATA} P.J. Mohr and B.N. Taylor, {\it Reviews of modern physics} {\bf 77} (2005), 1.

\bibitem{Tsai1975} W. Tsai and T. Erber, {\it Phys. Rev. D} {\bf 12} (1975), 1132.

\bibitem{Heyl1997_3} J.S. Heyl and L. Hernquist, {\it J. Phys. A: Math. Gen.} {\bf 30} (1997), 6485.

\bibitem{CasteloFerreira2008} P. Castelo Ferreira and J. Dias de Deus, {\it Eur. Phys. J. C} {\bf 54} (2008), 539.

\bibitem{Rikken2003} G.L.J.A. Rikken, and C. Rizzo, {\it Phys. Rev. A} {\bf 67} (2003), 015801.

\bibitem{BornWolf} Born \& Wolf, {\it Principles of optics}, $6^{th}$ ed., Pergamon Press.

\bibitem{Morley1898} E.W. Morley, H.T. Eddy, and D.C. Miller, {\it Bulletin Western Reserve University}, {\bf
1}  (1898), 50.

\bibitem{Watson1929} W. H. Watson, {\it Proc. Roy. Soc. London A} {\bf 125} (1929), 345.

\bibitem{Farr1932} C. C. Farr and C. J. Banwell, {\it Proc. Roy. Soc. London A} {\bf 137} (1932), 275.

\bibitem{Banwell1940} C. J. Banwell and C. C. Farr, {\it Proc. Roy. Soc. London A} {\bf 175} (1940), 1.

\bibitem{Erber1961} T. Erber, {\it Nature} {\bf 190} (1961), 25.

\bibitem{GrassiStrini1979} A.M. Grassi Strini, G. Strini and G. Tagliaferri, {\it Phys. Rev. D} {\bf 19} (1979), 2330.

\bibitem{Ni1991} W-T. Ni, K. Tsubono, N. Mio, K. Narihara, S-C. Chen, S-K. King and S-S. Pan, {\it Mod. Phys. Lett. A} {\bf 6} (1991), 3671.

\bibitem{Zavattini2009} G. Zavattini, and E. Calloni, {\it Eur. Phys. J. C} {\bf 62} (2009), 459.

\bibitem{Dobrich2009} B. D\"{o}brich and H. Gies, {\it EPL} {\bf 87} (2009), 21002.

\bibitem{LIGO} see : http://www.ligo.caltech.edu/

\bibitem{VIRGO} see : https://wwwcascina.virgo.infn.it/

\bibitem{Iacopini1979} E. Iacopini and E. Zavattini, {\it Phys. Lett. B} {\bf 85} (1979), 151.

\bibitem{Rizzo1997} C. Rizzo, A. Rizzo and D.M. Bishop, {\it Int. Rev. Phys. Chem.} {\bf 16} (1997), 81

\bibitem{CERN} http://public.web.cern.ch/public/

\bibitem{Iacopini1981} E. Iacopini, B. Smith, G. Stefanini and E. Zavattini, {\it Il Nuovo Cimento B} {\bf 61} (1981), 21.

\bibitem{BNL} http://www.bnl.gov/world/

\bibitem{Cameron1993} R. Cameron, G. Cantatore, A.C. Melissinos, G. Ruoso, Y. Semertzidis, H.J. Halama, D.M. Lazarus, A.G. Prodell, F. Nezrick, C. Rizzo and E. Zavattini, {\it Phys. Rev. D} {\bf 47} (1993), 3707.

\bibitem{Maiani1986} L. Maiani, R. Petronzio, and E. Zavattini, {\it Phys. Lett. B} {\bf 175} (1986), 359.

\bibitem{Gies2008} H. Gies, {\it J. Phys. A: Math. Theor.} {\bf 41} (2008), 164039.

\bibitem{LNL} http://www.lnl.infn.it/

\bibitem{Adler2007} S.L. Adler, {\it J. Phys. A: Math. Theor.} {\bf 40} (2007), F143.

\bibitem{Zavattini2008} E. Zavattini, G. Zavattini, G. Ruoso, G. Raiteri, E. Polacco, E. Milotti, V. Lozza, M. Karuza, U. Gastaldi, G. Di Domenico, F. Della Valle, R. Cimino, S. Carusotto, G. Cantatore, and M. Bregant, {\it Phys. Rev. D} {\bf 77} (2008), 032006.

\bibitem{NTHU} http://www.nthu.edu.tw/english/index.php

\bibitem{Ni1996} W-T Ni, {\it Chin. J. Phys.} {\bf 34} (1996), 962.

\bibitem{Mei2010} H-H. Mei, W-T. Ni, S-J Chen, and S-S Pan, {\it Mod. Phys. Lett. A} {\bf 25} (2010), 983.

\bibitem{Rizzo} C. Rizzo, {\it EPL} {\bf 41} (1998), 483

\bibitem{Battesti2008} R. Battesti, B. Pinto Da Souza, S. Batut, C. Robilliard, G. Bailly, C. Michel, M. Nardone, L. Pinard, O. Portugall, G. Tr\'enec, J-M. Mackowski, G.L.J.A. Rikken, J. Vigu\'e, and C. Rizzo,  {\it Eur. Phys. J. D} {\bf 46} (2008), 323.

\bibitem{BMVPRA} P. Berceau, M. Fouch\'e, R. Battesti, and C. Rizzo, {\it Phys. Rev. A} {\bf 85} (2012), 013837

\bibitem{Lee1995} S.A. Lee, W.M. Fairbank Jr., W.H. Toki, J.L. Hall, T.S. Jaffery, P. Colestock, V. Cupps, H. Kautzky, M. Kuchnir, and F. Nezrick, {\it Fermilab proposal} {\bf 877} (March 28 1995).

\bibitem{Fermilab} http://www.fnal.gov/

\bibitem{Pugnat2006} P. Pugnat, M. Kral, A. Siemko, L. Duvillaret, M. Finger, M. Finger, K.A. Meissner, D. Romanini, M. Sulc, and J. Zicha, {\it Czec J. Phys.} {\bf 56} (2006), C193.

\bibitem{Ferrara} http://www.fe.infn.it/

\bibitem{Zavattini2012} G. Zavattini, U. Gastaldi, R. Pengo, G. Ruoso, F. Della Valle, and E. Milotti, {\it Int. J. of Mod.
Phys. A} {\bf 27} (2012), 1260017.

\bibitem{Cantatore1991} G. Cantatore, F. Della Valle, E. Milotti, L. Dabrowski and C. Rizzo, {\it Phys. Lett. B} {\bf 265} (1991), 418.

\bibitem{Chen2012} Q.-F. Chen, A. Nevsky, and S. Schiller, {\it Appl. Phys. B} {\bf 107} (2012), 679.

\bibitem{Stedman1997} G.E. Stedman, {\it Rep. Prog. Phys.} {\bf 60} (1997), 615.

\bibitem{Hall2000} J.L. Hall, J. Ye, and L.S. Ma, {\it Phys. Rev. A} {\bf 62} (2000), 013815

\bibitem{Robilliard2010} G. Bailly, R. Thon, and C. Robilliard, {\it Rev. Sci. Instrumen.} {\bf 81}(2010), 033105.

\bibitem{Pelle2011} B. Pelle, H. Bitard, G. Bailly, and C. Robilliard, {\it Phys. Rev. Lett.} {\bf 106} (2011), 193003.

\bibitem{Aleksandrov1985} E.B. Aleksandrov, A.A. Ansel'm and A.N. Moskalev, {\it Soviet Physics-JETP} {\bf 62} (1985), 680.

\bibitem{Luiten2004} A.N. Luiten and J.C. Peterson, {\it Phys. Rev. A} {\bf 70} (2004), 033801.

\bibitem{Homma2011} K. Homma, D. Habs, and T. Tajima, {\it Appl. Phys. B} {\bf 104} (2011), 769.

\bibitem{PCFourier} M. Bass ed., {\it. Handbook of Optics}, McGraw-Hill, 1995.

\bibitem{Heinzl2006} T. Heinzl, B. Liesfeld, K-U Amthor, H. Schwoerer, R. Sauerbrey, and A. Wipf, {\it Opt. Com.} {\bf 267} (2006), 318.

\bibitem{DiPiazza2006} A. Di Piazza, K.Z. Hatsagortsyan, and C.H. Keitel, {\it Phys. Rev. Lett.} {\bf 97} (2006), 083603.

\bibitem{MarklundNP2010} M. Marklund, {\it Nature Photon.} {\bf 4} (2010), 72.

\bibitem{King2010NP} B. King, A. Di Piazza, and C.H. Keitel, {\it Nature Photon.} {\bf 4} (2010), 92.

\bibitem{Marx2011} B. Marx, I. Uschmann, S. H\"{o}fer, R. L\"{o}tzsch, O. Wehrhan, E. F\"{o}rster, M. Kaluza, T. St\"{o}hlker, H. Gies , C.
Detlefs, T. Roth, J. H\"{a}rtwig, and G.G. Paulus, {\it Optics Com.} {\bf 284} (2011), 915.

\bibitem{Baranga2011} A. Ben-Amar Baranga, R. Battesti, M. Fouch\'e, C. Rizzo, and G.L.J.A. Rikken, {\it EPL} {\bf 94} (2011), 44005

\bibitem{Rizzo2010} C. Rizzo, A. Dupays, R. Battesti, M. Fouch\'e, and G.L.J.A. Rikken, {\it EPL} {\bf 90} (2010), 64003.

\bibitem{Kalugin2001} N.G. Kalugin and G. Wagniere, {\it J. Opt. B: Quantum Semiclass. Opt.} {\bf 3} (2001), S189.

\bibitem{Adler1970} S.L. Adler, J.N. Bachall, G.C. Callan, and M.N. Rosenbluth, {\it Phys. Rev. Lett.} {\bf 25} (1970), 1061.

\bibitem{Adler1996} S.L. Adler and C. Schubert, {\it Phys. Rev. Lett.} {\bf 77} (1996), 1695.

\bibitem{Lee2003} R.N. Lee, A.L. Maslennikov, A.I. Milstein, V.M. Strakhovenko, and Yu.A. Tikhonov, {\it Phys. Rep.} {\bf 373} (2003), 213.

\bibitem{Akhmadaliev2002} Sh.Zh. Akhmadaliev, G.Ya. Kezerashvili, S.G. Klimenko, R.N. Lee, V.M. Malyshev, A.L. Maslennikov, A.M. Milov, A.I. Milstein, N.Yu. Muchnoi, A.I. Naumenkov, V.S. Panin, S.V. Peleganchuk, G.E. Pospelov, I.Ya. Protopopov, L.V. Romanov, A.G. Shamov, D.N. Shatilov, E.A. Simonov, V.M. Strakhovenko, and Yu.A. Tikhonov, {\it Phys. Rev. Lett.} {\bf 89} (2002), 061802.

\bibitem{BudInst} http://www.inp.nsk.su/index.en.shtml/

\bibitem{Affleck1987} I. Affleck and L. Kruglyak, {\it Phys. Rev. Lett.} {\bf 59} (1987), 1065.

\bibitem{DiPiazza2007PRA} A. Di Piazza, A.I. Milstein, and C.H. Keitel, {\it Phys. Rev. A} {\bf 76} (2007), 032103.

\bibitem{Raizen1990} M.G. Raizen and B. Rosenstein, {\it Phys. Rev. Lett.} {\bf 65} (1990), 2744.

\bibitem{Ford1990} G.W. Ford and D.G. Steel, {\it Phys. Rev. Lett.} {\bf 65} (1990), 2745.

\bibitem{Bethea1975} C.G. Bethea, {\it Appl. Opt. } {\bf 14} (1975), 2435.

\bibitem{Ding1992} Y.J. Ding and A.E. Kaplan, {\it Int. J. Nonlin. Opt. Phys.} {\bf 1} (1991), 51.

\bibitem{Kaplan2000} A.E. Kaplan and Y.J. Ding, {\it Phys. Rev. A} {\bf 62} (2000), 043805.

\bibitem{Miller2003} A.J. Miller, S. Woo Nam, J.M. Martinis, and A.V. Sergienko, {\it Appl Phys. Lett.} {\bf 83} (2003), 791.

\bibitem{Denisov2001} V.I. Denisov, I.P. Denisova, and S.I. Svertilov, {\it Doklady Phys.} {\bf 46} (2001), 705.

\bibitem{Dupays2005} A. Dupays, C. Robilliard, C. Rizzo, and G.F. Bignami, {\it Phys. Rev. Lett.} {\bf 94} (2005), 161101.

\bibitem{Jones1961} R.V. Jones, {\it Nature} {\bf 186} (1960), 706 ; R.V. Jones, {\it Proc. Roy. Soc. London A} {\bf 260} (1961), 47.

\bibitem{Soljacic2000} M. Soljacic and M. Segev, {\it Phys. Rev. A} {\bf 62} (2000), 043817.

\bibitem{Li2011} W. Li, J. Chen, G. Nouet, L-y. Chen, and X. Jiang, {\it Appl. Phys. Lett} {\bf 99} (2011), 051112.

\bibitem{Ferrando2007} A. Ferrando, H. Michinel, M. Seco, and D. Tommasini, {\it Phys. Rev. Lett.} {\bf 99} (2007), 150404.

\bibitem{Tommasini2010PRA} D. Tommasini, and H. Michinel, {\it Phys. Rev. A} {\bf 82} (2010), 011803(R).

\bibitem{King2010} B. King, A. Di Piazza, and C.H. Keitel, {\it Phys. Rev. A} {\bf 82} (2010), 032114.

\bibitem{Yu2011} G.Y, Kryuchkyan and K.Z. Hatsagortsyan, {\it Phys. Rev. Lett.} {\bf 107} (2011), 053604.

\bibitem{BraggS} L.D. Landau and E.M. Lifshitz, {\it Electrodynamics of Continuous Media}, Pergamon Press, 1984.

\bibitem{Karplus1951} R. Karplus and M. Neuman, {\it Phys. Rev.} {\bf 83} (1951), 776.

\bibitem{Liang2012} Y. Liang, and A. Czarnecki, {\it Can. J. Phys.} {\bf 90} (2012), 11.

\bibitem{PCM} The photon center of mass frame (also called photon center of momentum frame) is the inertial frame where the photon total energy is the smallest one.

\bibitem{DeTollis1964} B. De Tollis, {\it Nuovo Cimento} {\bf 32} (1964), 757 ; {\it Nuovo Cimento} {\bf 32} (1965), 1182.

\bibitem{Dicus1998} D.A. Dicus, C. Kao and W.W. Repko, {\it Phys. Rev. D} {\bf 57} (1998), 2443.

\bibitem{Vavilov1928} S. Vavilov, {\it Jour. Russ. Phys. Chem.} {\bf 60} (1928), 555.

\bibitem{Vavilov1930} S. Vavilov, {\it Phys. Rev.} {\bf 36} (1930), 1590.

\bibitem{Hughes1930} A.L. Hughes and G.E.M. Jauncey, {\it Phys. Rev.} {\bf 36} (1930), 773.

\bibitem{Harutyunian1964} V.M. Harutyunian, F.R. Harutyunian, K.A. Ispirian and V.A. Tumanian, {\it Phys. Lett.} {\bf 6} (1963), 175; V.M. Harutyunian, F.R. Harutyunian, K.A. Ispirian and V.A. Tumanian, {\it Soviet Physics-JETP} {\bf 18} (1964), 873.

\bibitem{Rosen1965} G. Rosen and F.C. Whitmore, {\it Phys. Rev.} {\bf 137} (1965), 1357.

\bibitem{Moulin1996} F. Moulin, D. Bernard, and F. Amiranoff, {\it Z. Phys. C} {\bf 72} (1996), 607.

\bibitem{Dewar1974} R.L. Dewar, {\it Phys. Rev. A} {\bf 10} (1974), 2107.

\bibitem{Grynberg1990} G. Grynberg and J-Y. Courtois, {\it C. R. Acad. Sci. Paris} {\bf 311} (1990), 1149.

\bibitem{Bernard2000} D. Bernard, F. Moulin, F. Amiranoff, A. Braun, J.P. Chambaret, G. Darpentigny, G. Grillon, S. Ranc, and F. Perrone, {\it Eur. Phys. J. D} {\bf 10} (2000), 141.

\bibitem{Moulin1999} F. Moulin, and D. Bernard, {\it Opt. Com.} {\bf 164} (1999), 137.

\bibitem{Brodin2001} G. Brodin, M. Marklund, and L. Stenflo, {\it Phys. Rev. Lett.} {\bf 87} (2001), 171801.

\bibitem{Lundin2006} J. Lundin, M. Marklund, E. Lundstr\"{o}m, G. Brodin, J. Collier, R. Bingham, J.T. Mendo\c{c}a, P. Norreys, {\it Phys. Rev. A} {\bf 74} (2006), 043821.

\bibitem{Lundstrom2006} E. Lundstr\"{o}m, G. Brodin, J. Lundin, M. Marklund,  R. Bingham, J. Collier,  J.T. Mendo\c{c}a, P. Norreys, {\it Phys. Rev. Lett.} {\bf 96} (2006), 083602.

\bibitem{Fedotov2007} A.M. Fedotov, and N.B. Narozhny, {\it Phys. Lett. A} {\bf 362} (2007), 1.

\bibitem{Tommasini2008} D. Tommasini, A. Ferrando, H. Michinel, and M. Seco, {\it Phys. Rev. A} {\bf 77} (2008), 042101.

\bibitem{Tommasini2009} D. Tommasini, A. Ferrando, H. Michinel, and M. Seco, {\it Journal of high energy physics  } {\bf 11}  (2009), 43.

\bibitem{Tommasini2010} D. Tommasini, and H. Michinel, {\it Phys. Rev. A} {\bf 82} (2010), 011803(R).

\bibitem{Marklund2009} M. Marklund and J. Lundin, {\it Eur. Phys. J. D} {\bf 55} (2009), 319.

\bibitem{Heinzl2009} T. Heinzl and A. Ilderton, {\it Eur. Phys. J. D} {\bf 55} (2009), 359.

\bibitem{Marklund2006RMP} M. Marklund and P.K. Shukla, {\it Rev. Mod. Phys.} {\bf 78} (2006), 591.

\bibitem{Haissinski2006} J. Ha\"{i}ssinski, S. Dagoret, M. Urban, and F. Zomer, {\it Physica Scripta} {\bf 74} (2006), 678.

\bibitem{Bregant2008} M. Bregant, G. Cantatore, S. Carusotto, R. Cimino, F. Della Valle, G. Di Domenico, U. gastaldi, M. Karuza, V. Lozza, E. Milotti, E. Polacco, G. Raiteri, G. Ruoso, E. Zavattini, and G. Zavattini, {\it Phys. Rev. D} {\bf 78} (2008), 032006.

\bibitem{Klein1964} J.J. Klein and B.P. Nigam, {\it Phys. Rev.} {\bf 136} (1964), B1540.

\bibitem{Dunne2009} G.V. Dunne, H. Gies, and R. Schutzhold, {\it Phys. Rev. D} {\bf 80} (2009), 111301(R).

\bibitem{Labun2011} L. Labun and J. Rafelski, {\it Phys. Rev. D} {\bf 84} (2011), 033003.

\bibitem{Delbruck1933} M. Delbr\"{u}ck, {\it Z. Physik} {\bf 84} (1933), 144.

\bibitem{Meitner1933} L. Meitner, H. K\"{o}ster, {\it Z. Physik} {\bf 84} (1933), 137.

\bibitem{Rohrlich1952} F. Rohrlich and R.L. Gluckstern, {\it Phys. Rev.} {\bf 86} (1952), 1.

\bibitem{Milstein1994} A.I. Milstein, and M. Schumacher, {\it Phys. Reports} {\bf 243} (1994), 183.

\bibitem{Schumacher1999} M. Schumacher, {\it Rad. Phys. Chem.} {\bf 56} (1999), 101.

\bibitem{Schumacher1975} M. Schumacher, I. Borchert, F. Smend, and P. Rullhusen,  {\it Phys. Letters B} {\bf 59} (1975), 134.

\bibitem{Akhmadalev1998} S.Z. Akhmadaliev, et al., {\it Phys. Rev. C} {\bf 58} (1998), 2844.

\bibitem{Harding2006} A.K. Harding and D. Lai, {\it Rep. Prog. Phys.} {\bf 69} (2006), 2631.

\bibitem{Koester2002} D. Koester, {\it The Astron. Astrophys. Rev.} {\bf 11} (2002), 33.

\bibitem{HeylCargese} J.S. Heyl, {\it Proc. QED2012}, (2012) 04002, open access at http://www.iesc-proceedings.org/.

\bibitem{Cuesta2006} H.J.M. Cuesta, J.M. Salim, and J.A.D. Pacheco, {\it Int. J. Mod. Phys. A} {\bf 21} (2006), 43.

\bibitem{Bartelmann2001} M. Bartelmann and P. Schneider, {\it Phys. Rep.} {\bf 340} (2001), 291.

\bibitem{Xue2003} S-S. Xue, {\it Phys. Rev. D} {\bf 68} (2003), 013004.

\bibitem{BHs} P.C.W. Davies, {\it Rep. Prog. Phys.} {\bf 41} (1978), 1313.

\bibitem{Treves1999} A. Treves and R. Turolla, {\it ApJ} {\bf 517} (1999) 396.

\bibitem{Dupays2008} A. Dupays, C. Rizzo, D. Bakalov and G.F. Bignami, {\it EPL} {\bf 82} (2008), 69002.

\bibitem{Dupays2012} A. Dupays, C. Rizzo, and G.F. Bignami, {\it EPL} {\bf 98} (2012), 49001.

\bibitem{Heyl2002} J.S. Heyl, and N.J. Shaviv, {\it Phys. Rev. D} {\bf 66} (2002), 023002.

\bibitem{Lai2003} D. Lai and W.C.G. Ho, {\it Phys. Rev. Lett.} {\bf 91} (2003), 071101.

\bibitem{Potekhin2004} A.Y. Potekhin, D. Lai, G. Chabrier, and W.C.G. Ho, {\it ApJ.} {\bf 612} (2004), 1034.

\bibitem{Kerkwijk2004} M.H. van Kerkwijk, D.L. Kaplan, M. Durant, S.R. Kulkarni, and F. Paerels, {\it ApJ.} {\bf 608} (2004), 432.

\bibitem{Shannon2006} R.M. Shannon and J.S. Heyl, {\it Mon. Not. R. Astron. Soc.} {\bf 368} (2006), 1377.

\bibitem{Gnedin2006} Yu.N. Gnedin, N.V. Borisov, V.M. larionov, T.M. Natsvlishvili, M.Yu. Piotrovich, and A.A. Arkharov, {\it Astron. Rep.} {\bf 50} (2006), 553.

\bibitem{vanAdelsberg2008} M. van Adelsberg and D. Lai, {\it AIP Conf. Proc.} {\bf 968} (2008), 137.

\bibitem{Wang2009} C. Wang and D. Lai, {\it Mon. Not. R. Astron. Soc.} {\bf 398} (2009), 515.

\bibitem{Heyl1999} J.H. Heyl, and L. Hernquist, {\it Phys. Rev. D} {\bf 59} (1999), 45005.

\bibitem{DiPiazza2007} A. Di Piazza, K.Z. Hatsagortsyan, and C.H. Keitel, {\it Phys. Plas.} {\bf 14} (2007), 032101.

\bibitem{Brodin2007PRL} G. Brodin, M. Marklund, B. Eliasson, and P.K. Shukla, {\it Phys. Rev. Lett.} {\bf 98} (2007), 125001.

\bibitem{Mazur2011} D. Mazur and J.S. Heyl, {\it Mon. Not. R. Astron. Soc.} {\bf 412} (2011), 1381.

\bibitem{Baring2001} M.G. Baring and A.K. Harding, {\it ApJ} {\bf 547} (2001), 929.

\bibitem{GEMS} http://heasarc.gsfc.nasa.gov/docs/gems/

\bibitem{Scharnhorst1990} K. Scharnhorst, {\it Phys. Lett. B} {\bf 236} (1990), 354.

\bibitem{Casimir1948} H.B.G. Casimir, {\it Proc. Kon. Nederland. Akad. Wetensch.} {\bf B51} (1948), 793.

\bibitem{Milonni1990} P.W. Milonni and K. Svozil, {\it Phys. Lett. B} {\bf 248} (1990), 437.

\bibitem{Dittrich1998} W. Dittrich and H. Gies, {\it Phys. Rev. D} {\bf 58} (1998), 025004.

\bibitem{Brodin2003} G. Brodin, L. Stenflo, D. Anderson, M. Lisak, M. Marklund, and P. Johannisson, {\it Phys. Lett. A} {\bf 306} (2003), 206.

\bibitem{Marklund2004} M. Marklund, B. Eliasson, and P.K. Shukla, {\it JETP Lett.} {\bf 79} (2004), 262.

\bibitem{Shukla2004PRL} P.K. Shukla and B. Eliasson, {\it Phys. Rev. Lett.} {\bf 92} (2004), 073601.

\bibitem{Marklund2005} M. Marklund, P.K. Shukla, and B. Eliasson, {\it EPL} {\bf 70} (2005), 327.

\bibitem{Mendonca2006} J.T. Mendonca, M. Marklund, P.K. Shukla, and G. Brodin, {\it Phys. Lett. A} {\bf 359} (2006), 700.

\bibitem{Aubert2008} B. Aubert, et al. (The BABAR Collaboration), {\it Phys. Rev. Lett.} {\bf 101} (2008), 021801.

\bibitem{Pinto2006} B. Pinto Da Souza, R. Battesti, C. Robilliard, and C. Rizzo,  {\it Eur. Phys. J. D} {\bf 40} (2006), 445.

\bibitem{Millo2009} R. Millo and P. Faccioli, {\it Phys. Rev. D} {\bf 79} (2009), 65020.

\bibitem{Gabrielse2006} G. Gabrielse, D. Hanneke, T. Kinoshita, M. Nio, and B. Odom, {\it Phys. Rev. Lett.} {\bf 97} (2006), 030802 ; {\it Phys. Rev. Lett.} {\bf 99} (2007), 039902.

\bibitem{Heinzl2012} T. Heinzl, {\it Int. J. Mod. Phys. A} {\bf 27} (2012), 1260010.

\bibitem{DiPiazza2012} A. Di Piazza, C. M\"{u}ller, K.Z. Hatsagortsyan and C.H. Keitel, {\it Rev. Mod. Phys.} {\bf 84} (2012), 1177.

\end{thebibliography}
\end{document}